\documentclass{aa}  

\usepackage{graphicx}
\usepackage{subcaption}
\graphicspath{{./pictures/}}

\usepackage{txfonts}

\begin{document}

   \title{Structure of X-ray emitting jets close to the launching site: 
   from embedded to disk-bearing sources}

   \author{S. Ustamujic
          \inst{1}
          \and
          S. Orlando
          \inst{2}
          \and
          R. Bonito
          \inst{2,3}
          \and
          M. Miceli
          \inst{3,2}
          \and
          A. I. G\'omez de Castro
          \inst{1}       
          }

   \institute{S. D. Astronom\'ia y Geodesia, Facultad de Ciencias
   			  Matem\'aticas, Universidad Complutense de Madrid, 28040 Madrid, 
   			  Spain\\
              \email{sustamuj@ucm.es}
         \and
             INAF-Osservatorio Astronomico di Palermo, Piazza del Parlamento 1,
             90134 Palermo, Italy\\
         \and
             Dipartimento di Fisica e Chimica, Universit\`a di Palermo, Via 
             Archirafi 36, 90123 Palermo, Italy \\
             }

   \date{Received 1 December 2017; accepted 15 March 2018}

 
  \abstract
   {Several observations of stellar jets show evidence of X-ray emitting 
   shocks close to the launching site. In some cases, the shocked features 
   appear to be stationary, also for Young Stellar Objects (YSOs) at different 
   stages of evolution. We study the case of HH~154, the jet originating 
   from the embedded binary Class 0/I protostar IRS 5, and the case of the 
   jet associated to DG~Tau, a more evolved Class II disk-bearing source 
   or Classical T Tauri star (CTTS), both located in the Taurus star-forming 
   region.}
   {We aim at investigating the effect of perturbations in X-ray emitting 
   stationary shocks in stellar jets; the stability and detectability in X-rays 
   of these shocks; and explore the differences in jets from Class 0 to 
   Class II sources.}
   {We performed a set of 2.5-dimensional magnetohydrodynamic 
   numerical simulations that modelled supersonic jets ramming into a 
   magnetized medium. The jet is formed by two components: a continously 
   driven component that forms a quasi-stationary shock at the base of the 
   jet; and a pulsed component constituted by blobs perturbing the shock. 
   We explored different parameters for both components. 
   We studied two cases: a jet less dense than the ambient medium (light jet),
   representing the case of HH~154; and a jet denser than the ambient 
   (heavy jet), associated with DG~Tau. We synthesized the count rate from 
   the simulations and compared with available \textit{Chandra} observations.}
   {Our model explains the formation of X-ray emitting quasi-stationary 
   shocks observed at the base of jets in a natural way, being able to 
   reproduce the observed jet properties at different evolutionary phases
   (in particular, for HH~154 and DG~Tau). The jet is collimated by the 
   magnetic field forming a quasi-stationary shock at the base which 
   emits in X-rays even when perturbations formed by a train of blobs are 
   present. We found similar collimation mechanisms dominating in both 
   heavy and light jets.} 
   {We derived the pysical parameters that can give rise to X-ray emission 
   consistent with observations of HH~154 and DG~Tau. We have also 
   performed a wide exploration of the parameter space characterizing the 
   model; this can represent a useful tool to study and diagnose the physical 
   properties of YSO jets over a broad range of physical conditions, from 
   embedded to disk-bearing sources.
   We have shown that luminosity did not change significantly in variable jet 
   models for the range of parameters explored.
   Finally, we provided an estimation of the maximum perturbations that can 
   be present in HH~154 and DG~Tau taking into account the available X-ray 
   observations.}


   \maketitle


\section{Introduction}
\label{sec:intro}

   Young Stellar Objects (YSOs) are stars at their early stages of evolution 
   which are characterized by the amount of circumstellar material and its 
   interaction with the forming star.
   The principal phases of YSO evolution comprises protostars, Classical 
   T Tauri (CTT) stars, and weak-lined T Tauri (WTT) stars. Their 
   evolutionary phase is generally classified by their infrared-millimeter 
   spectral energy distributions from Class~0 to Class~III objects 
   \citep[see][]{lad87,and94}.
   Class~0 sources are young infalling protostars with massive, cold and large 
   extent envelopes that collapse toward the central regions. Class~I sources 
   are most evolved protostars but still surrounded by an optically thick 
   envelope. Subsequently, when the surrounding envelope disperses accreted 
   onto the disk or star, or dispersed by the outflow, we refer to them as 
   Class~II objects. In this phase the star is optically visible as a CTT star 
   which possess an extensive disk, and most of its complex phenomenology 
   can be modeled as a star interacting with a circumstellar accretion disk 
   \citep{sto94,sto96,rom11,orl11,zan13}. 
   Finally, when only the star with little or no accretion disk is left, we classify 
   them as Class III sources or WTT stars.
   For a complete description of the various protostellar and stellar phases 
   see \citet{fei99}.

   A variety of mass ejection phenomena occur during these first stages of star 
   evolution that are tightly connected to the accretion process (for an overview, 
   see \citealt{fra14}). Jets are always present during the accretion process, 
   and they are believed to carry away angular momentum \citep{bac02,cof07} 
   allowing the material in the outer disk to be transported to the inner disk and 
   continue accreting towards the central object. 
   The strong correlation between ejection and accretion found in 
   pre-main sequence stars \citep{cab90,har95} suggests that time variability 
   in the accreting disk produces variability in the associated jet.
   Models suggest that jets are launched and collimated by a symbiosis of 
   accretion, rotation and magnetic mechanisms (for a review, see \citealt{pud07}). 
   In light of the accretion-powered extended disk wind model (initially proposed 
   by \citealt{pud83,pud86}) outflows are driven magneto-centrifugally from the 
   inner portion of accretion disks and dense plasma from the disk is collimated 
   into jets \citep{fer06}.
  
   The general consensus is that magnetic fields play a crucial role in 
   launching, collimating and stabilizing the plasma of jets in young stellar objects.
   This idea was investigated both by measurements of multiple observational 
   data (see \citealt{cab07_LNP} for a review) and by extensive numerical 
   simulations. It is revealed that magnetohydrodynamic (MHD) 
   self-collimation and acceleration is most likely required at all stages of star 
   formation \citep{cab07}, appearing as the most effective mechanism able to 
   reproduce the observed jet properties at all evolutionary phases
   \citep{cab07_LNP}. Recently, \citet{ust16} found that the magnetic field plays 
   a major role in collimating the plasma at the base of the jet and in producing 
   there a stationary X-ray emitting shock. This idea was also corroborated by 
   scaled laboratory experiments \citep[see][]{cia09,alb14}.
  
   Usually jets are detected by their interaction with the sorrounding medium 
   forming the so-called Herbig-Haro (HH) objects, that have been 
   observed in a wide range of evolutionary stages (from Class 0 to Class II) 
   and in different wavelength bands (see the review of \citealt{rei01}). The 
   knotty structure observed along the jet axis is interpreted as the 
   consequence of the pulsing nature of the ejection of material by the star 
   (e.g. \citealt{rag90,rag07}; \citealt{col06}; \citealt{bon10b,bon10a}; 
   and references therein). The variable ejection jet model was succesfully 
   applied to several HH objects reproducing well the observed structures: 
   HH~34 \citep{rag98}, DG~Tau \citep{rag01}, HH~111 \citep{mas02}, 
   HH~32 \citep{rag04}, HH~154 \citep{bon10a}, and HH 444 \citep{rag10}.
  
   X-ray observations showed evidence of faint X-ray emitting sources 
   forming within the jet (e.g. 
   \citealt{pra01,fav02,bal03,pra04,tsu04,gud05,ste09}). These observations 
   were investigated through HD models which have shown that they are 
   consistent with the production of strong shocks that heat the plasma up to 
   temperatures of a few million degrees emitting in X-rays 
   \citep{bon07,bon10b,bon10a}. 
   Moreover, in the best-studied X-ray jets (HH~154, \citealt{fav06}; DG~Tau, 
   \citealt{gud05}), both located in the Taurus molecular complex at distance 
   of $\approx 140$\,pc, the shocked features appeared to be stationary and
   located close to the base of the jet.  
   In HH~154 the jet originates from the deeply embedded binary Class 0/I 
   protostar IRS 5 in the L1551 star-forming region \citep{rod98}. On the other 
   hand, in DG~Tau the jet originates from a more evolved Class II 
   disk-bearing source (CTTS) (see \citealt{eis98} for a detailed description).
   
   In our previous study, we investigated the formation of X-ray emitting 
   stationary shocks in magnetized protostellar jets through 2.5D MHD 
   simulations \citep{ust16}. We showed that a continuously driven stellar 
   jet forms a stationary X-ray emitting shock at the base with physical 
   parameters in good agreement with observations. According to the YSO jets
   phenomenology described in this section, we do not expect a continuous 
   flow propagating but rather a variable perturbed plasma 
   (see \citealt{ste15,ste17} for a brief description of the variability observed at YSOs). 
   Here we aim at investigating the effect of perturbations in the X-ray emitting 
   stationary shocks formed at the base of stellar jets described in \citet{ust16}.
   We propose a MHD model composed by two components: a continously driven 
   component that forms a stationary shock at the base of the jet (see \citealt{ust16}); 
   and a pulsed component formed by blobs variable in density, velocity and radius. 
   We apply our model to the X-ray jets of HH~154 and DG~Tau, and we compare 
   the results with observations via the count rate synthesized from the simulations.
   The distinct stage of evolution of the two objects selected (HH~154 and DG~Tau) 
   allows us to explore the possible various mechanisms present at different stages 
   of evolution in YSOs. 
   These studies are important to better understand the evolution of YSOs and 
   the structure of HH objects, and may give some insight into the still debated 
   jet collimation and acceleration mechanisms.
  
   The organization of this paper is as follows. In Section 2, we describe the MHD 
   model and the numerical setup and parameters. The results of our numerical 
   simulations for the two cases described are reported in Section 3. Finally, the 
   discussion and conclusions are presented in Section 4.


\section{The model}

   The  model describes the propagation of a stellar jet through an initially 
   isothermal and homogeneous magnetized medium. 
   We assume that the fluid is fully ionized and that it can be regarded as a 
   perfect gas with a ratio of specific heats $\gamma = 5/3$.\footnote{We 
   verified the assumptions used in this paper as described in \citet{bon07}.}
   
   The system is described by the time-dependent MHD equations extended with 
   radiative losses from optically thin plasma. We neglect the effect of thermal 
   conduction as it has been shown in \citet{ust16} that the evolution of the 
   post-shock plasma is dominated by the radiative cooling, whereas the thermal 
   conduction slightly affects the structure of the shock. The time-dependent 
   MHD equations written in non-dimensional conservative form are
   \begin{equation}
      \frac{\partial \rho}{\partial t} 
      + \nabla \cdot (\rho \boldsymbol{u}) = 0,
   \end{equation}
   \begin{equation}
      \frac{\partial (\rho \boldsymbol{u}) }{\partial t} + \nabla \cdot 
      (\rho\boldsymbol{u}\boldsymbol{u} - \boldsymbol{B}\boldsymbol{B}) 
      + \nabla P_{\mathrm{t}} = 0,
   \end{equation}
   \begin{equation}
      \frac{\partial (\rho E) }{\partial t}
      + \nabla \cdot [\boldsymbol{u} (\rho E+P_{\mathrm{t}})
      - \boldsymbol{B}(\boldsymbol{u}\boldsymbol{B})] = 
      - n_{\mathrm{e}} n_{\mathrm{H}} \Lambda (T),
   \label{eq.energy}
   \end{equation}
   \begin{equation}
      \frac{\partial \boldsymbol{B}}{\partial t} + \nabla \cdot
      (\boldsymbol{u}\boldsymbol{B}-\boldsymbol{B}\boldsymbol{u}) = 0,
   \end{equation}
   where
   \begin{equation}
      P_{\mathrm{t}} = P + \frac{B^2}{2}, 
      \qquad
      E = \epsilon + \frac{1}{2} u^2 + \frac{1}{2} \frac{B^2}{\rho},
   \end{equation}
   are respectively the total pressure and the total gas energy per unit mass 
   (internal energy $\epsilon$, kinetic energy, and magnetic energy per unit mass), 
   $t$ is the time, $\rho = \mu m_{\mathrm{H}} n_{\mathrm{H}}$ is the mass density, 
   $\mu = 1.29$ is the mean atomic mass (assuming solar abundances; 
   \citealt{and89}), $m_{\mathrm{H}}$ is the mass of the hydrogen atom, 
   $n_{\mathrm{H}}$ is the hydrogen number density, $\boldsymbol{u}$ is 
   the gas velocity, $\boldsymbol{B}$ is the magnetic field, $T$ is the 
   temperature, and $\Lambda (T)$ represents the optically thin radiative 
   losses per unit emission measure derived with the PINTofALE spectral 
   code \citep{kas00} and with the APED V1.3 atomic line database 
   \citep{smi01}, assuming solar metal abundances as before (as deduced 
   from X-ray observations of CTTSs; \citealt{tel07}). We use the ideal gas 
   law, $P = (\gamma-1)\rho\epsilon$.

   The calculations were performed using PLUTO \citep{mig07}, a modular 
   Godunov-type code for astrophysical plasmas. The code provides a 
   multiphysics, multialgorithm modular environment particularly oriented 
   towards the treatment of astrophysical flows in the presence of 
   discontinuities as in our case. The code was designed to make efficient 
   use of massive parallel computers using the message-passing interface 
   (MPI) library for interprocessor communications. The MHD equations are 
   solved using the MHD module available in PLUTO, configured to compute 
   intercell fluxes with the Harten-Lax-Van Leer Discontinuities (HLLD)
   approximate Riemann solver, while second order in time is achieved using 
   a Runge-Kutta scheme. The evolution of the magnetic field is carried out 
   adopting the constrained transport approach \citep{bal99} that maintains 
   the solenoidal condition ($\nabla\cdot B = 0$) at machine accuracy. 
   PLUTO includes optically thin radiative losses in a fractional step 
   formalism \citep{mig07}, which preserves the second time accuracy, as 
   the advection and source steps are at least of the second-order accurate; 
   the radiative losses ($\Lambda$ values) are computed at the temperature 
   of interest using a table lookup/interpolation method.

\subsection{Numerical setup}
\label{sec:num}

   We adopt a 2.5D cylindrical ($r$, $z$) coordinate system, assuming 
   axisymmetry. We consider the jet axis coincident with the $z$-axis. The 
   computational grid size ranges from $\approx200$ AU to $\approx600$ AU 
   in the $r$ direction and from $\approx1200$ AU to $\approx1600$ AU in the 
   $z$ direction, depending on the model parameters. We follow the evolution 
   of the system for at least 90-100 years. These dimensions and times are 
   comparable with those of the observations and are chosen so that we are 
   able to follow the formation and evolution of the shock diamond formed at 
   the base of the jet.
   
   We consider two different sets of parameters for our numerical setup 
   corresponding with the two different cases studied: 
   (1) light jet scenario (a jet initially less dense than the ambient medium) 
   representing the case of HH~154 \citep{bon04,bon08}, and 
   named by the letters ``LJ''; 
   (2) heavy jet scenario (a jet initially denser than the ambient medium) 
   representing the case of the jet associated to DG~Tau 
   \citep{gud05,gud08}, and named by the letters ``HJ''. 
   More specifically, the HH~154 case is well described by the LJ 
   scenario because the jet originates from a deeply embedded binary 
   Class 0/I protostar (\citealt{bon04,bon08} demonstrated that only the 
   scenario of a light jet can reproduce the HH~154 jet observations), 
   while the DG~Tau case comes from a more evolved Class II source, 
   typically described by the HJ scenario \citep{gud05,gud08}. 
    
   We define an initially isothermal and homogeneous magnetized static medium.
   The initial temperature and density of the ambient are fixed to 
   $T_{\mathrm{a}}=10\,$K and $n_{\mathrm{a}}=5000\,$cm$^{-3}$ 
   respectively in the LJ case. For the HJ case the values are $T_{\mathrm{a}}=100\,$K 
   and $n_{\mathrm{a}}=100\,$cm$^{-3}$ respectively. These values are selected 
   in order to find a jet-to-ambient density ratio, $\rho_{\mathrm{j}}/\rho_{\mathrm{a}}$, 
   of $\sim 0.1$ in the LJ scenario, and $\sim 10$ in the HJ case.
   We define a jet, injected into the domain at $z=0$, embedded in an initially axial ($z$) 
   and uniform magnetic field of strength $B_{\mathrm{z}}=$ 4~mG. 
   This value for the magnetic field strength was chosen according to the 
   values investigated in \cite{ust16}. It is also consistent with that estimated by 
   \cite{bon11} at the exit of a magnetic nozzle close to the base of the jet, 
   namely $B= 5$~mG, and by \cite{sch11}, who find $B\approx~6$ mG.  
   \cite{bal03} infered values around $B=1-4$~mG, in the context of shocks 
   associated with jet collimation dominated by static magnetic pressure. 
   An external magnetic field has to be defined as previous works revealed 
   that the ambient pressure alone is not sufficient to confine the jets and that 
   MHD self-collimation is most likely required \citep{cab07,bon11,ust16}.
   In some simulations we consider the plasma of the jet characterized by an 
   angular velocity corresponding to maximum linear rotational velocitiy of 
   $\varv_{\varphi,\mathrm{max}}=150$~km~s$^{-1}$ at the lower boundary. 
   In these cases a toroidal magnetic field component arises and 
   the magnetic field lines are twisted obtaining a helical shaped field 
   (see Fig. 7 in \citealt{ust16}). With increasing $\varv_{\varphi,\mathrm{max}}$ the 
   shock diamond is a brighter X-ray source with higher X-ray luminosity 
   (see \citealt{ust16} for more details).   

   The jet velocity and density are defined at the z-lower boundary in 
   order to have a mass ejection rate of $\approx 10^{-8}\,M_{\odot}$~yr$^{-1}$. 
   The jet is formed by two components: a continously driven component 
   that forms a stationary shock at the base of the jet (see \citealt{ust16}); 
   and a pulsed component starting after the stationary shock is formed, 
   in order to study the effect of perturbations in the stationarity of the shock. 
   As for the pulsed component, we followed \citet{bon10a} and assumed 
   that it consists by a train of blobs characterized by a density contrast 
   and/or a velocity contrast with respect to the continuous driven component. 
   The blobs represent variations in the mass ejection rate that can be due 
   to changes in the physical conditions of the jet launching site. In fact, 
   as largely debated in the literature, the dynamic and energetic phenomena 
   resulting from the star-disk interaction are expected to produce variations 
   in the physical paramaters of the jet. The pulsed component is introduced 
   after the jet reaches a quasi-stationary condition, namely 70 years in 
   case (A) and 40 years in case (B). We follow the evolution of the pulsed 
   jet for approximately 50 years. Following \citet{bon10a}, we define a blob 
   every 2 years with a duration of 0.5~yr.
   The initial radius and velocity of the jet continuous component are 30 AU and
   $\varv_{\mathrm{j}} = 500\,$km~s$^{-1}$ respectively in all the cases, while 
   different values are explored for the radius and velocities of the pulsed 
   component (see Sect.~\ref{sec:description} and Table~\ref{parameters}).
   In all the cases we use steepness profiles for the shear layer, adjusted so 
   as to achieve a smooth transition of the kinetic energy at the interface 
   between the jet and the ambient medium and to avoid possible numerical 
   artifacts that may develop there \citep{bon07}.
   The jet velocity at the lower boundary is oriented along the $z$-axis for both 
   pulsed and continuous component, coincident with the jet axis, and has a 
   radial profile of the form
   \begin{equation}
   V(r) = \dfrac{V_0}{\nu \, \mathrm{cosh}(r/r_\mathrm{j})^{\omega}-(\nu-1)},
   \end{equation}
   where $V_0$ is the on-axis velocity, $\nu$ is the ambient to jet density 
   ratio, $r_\mathrm{j}$ is the initial jet radius, and $\omega = 4$ is the 
   steepness parameter for the shear layer, adjusted so as to achieve a 
   smooth transition of the kinetic energy at the interface between the jet 
   and the ambient medium \citep{bon07}.
   The density variation in the radial direction is given by
   \begin{equation}
   \rho(r) = \rho_\mathrm{j} \left( \nu - 
   \dfrac{\nu-1}{\mathrm{cosh}(r/r_\mathrm{j})^{\omega}} \right),
   \end{equation}
   where $\rho_j$ is the jet density \citep{bod94}.

   The mesh is uniformly spaced along the two directions, giving a spatial 
   resolution of $1$~AU in the light jet scenario and $0.5$ in the heavy jet 
   scenario (corresponding to 30 and 60 cells across the initial jet diameter 
   respectively). We performed a convergence test to find the spatial 
   resolution needed to model the physics involved and to resolve the 
   X-ray emitting features. The test consisted in considering the setup for a 
   reference case and performing few simulations with increasing spatial 
   resolution. We found that by increasing the resolution adopted in our study 
   by a factor of 2, the results change by no more than 1\%. The domain was 
   chosen according to the physical scales of typical jets from young stars. 
   The adopted resolution is higher than that achieved by current instruments 
   used for the observations of jets, as HST in the optical band and 
   \textit{Chandra} in X-rays. In comparison, the \textit{Chandra} 
   resolution corresponds to $\sim 60$~AU at the distance of HH~154 and 
   DG~Tau in Taurus ($\sim 140$~pc).

   Axisymmetric boundary conditions are imposed along the jet axis 
   (at the left boundary for $r = 0$) in all the cases. At the lower 
   boundary (namely for $z = 0$), inflow boundary conditions (according to 
   the jet parameters given in Section~\ref{sec:description}) are imposed for 
   $r \leq r_{\mathrm{j}}$ (where $r_{\mathrm{j}}$ is the jet radius at the 
   lower boundary); for $r \geq r_{\mathrm{j}}$ we imposed boundary 
   conditions fixed to the ambient values prescribed at the initial conditions 
   (see the beginning of this section). Finally, outflow boundary conditions 
   are assumed elsewhere.

\subsection{Parameters}
\label{sec:description}
 
   Our model solutions depend upon a number of physical parameters, such 
   as the jet temperature, density, velocity (including a possible rotational 
   velocity $\varv_{\varphi}$) and radius. 
   In all the cases explored, we defined a jet density, velocity and radius in order 
   to preserve a mass ejection rate of the order of $10^{-8}\,M_{\odot}$ yr$^{-1}$. 
   Typical outflow rates are found to be 
   between $10^{-7}$ and $10^{-9}$ $M_{\odot}$ yr$^{-1}$ for jets from 
   low-mass CTTSs \citep{cab07,pod11}.
   We calculate the mass loss rate as 
   $\dot M_{\mathrm{j}} = \int \rho_\mathrm{j} \varv_\mathrm{j}\,\mathrm{d}A$, 
   where $\rho_\mathrm{j}$ and $\varv_\mathrm{j}$ are the mass density and jet 
   velocity, respectively, and $\mathrm{d}A$ is the cross-sectional area of 
   the incoming jet plasma.
   
   The jet temperature at the lower boundary is assumed to be 
   $T_{\mathrm{j}}=1-3\cdot10^6\,$K in order to obtain, as a result 
   of the jet expansion, values of $\approx 10^4-10^5$~K before the formation 
   of the shock diamond\footnote{See Fig. 3 in \citet{ust16}.}, in agreement 
   with the observational evidence that jets emit mainly in the optical/UV 
   band\footnote{See \citet{bon08} and Fig. 4 in \citet{bon10a}.}.
   The values used for the model are in good agreement with the observations 
   \citep{fri98,fav02,gud08}.
   The particle number density of the jet continuous component, $n_{\mathrm{j}}$, 
   ranges between $1-3\cdot10^{4}\,$cm$^{-3}$ at the lower boundary. When the 
   jet is injected into the domain the plasma expands and then is collimated by 
   the magnetic field. During this process the density decreases, leading to 
   pre-shock densities of the order of $10^{2}-10^{3}$~cm$^{-3}$, consistent 
   with those inferred by \cite{bal03}. 
   The exploration of the parameter space mainly focuses on the density,
   velocity and radius of the blobs composing the pulsed jet, that are defined as 
   function of the jet continuous component. 
   For the initial blob-to-jet particle number density ratio, 
   $\chi_{\mathrm{b}} = n_{\mathrm{b}}/n_{\mathrm{j}}$, 
   we explore values of $1.5$, $3$ and $10$.
   The blob velocity (also defined at the lower boundary) is the same
   as those of the continuous component in most cases, namely 
   $\varv_{\mathrm{b}} = 500$~km~s$^{-1}$, and takes random velocity with values 
   between 300 and 700~km~s$^{-1}$ in the rest of models 
   (see Table~\ref{parameters}).
   The initial blob-to-jet radius ratio, $R_{\mathrm{b}}=r_{\mathrm{b}}/r_{\mathrm{j}}$, 
   is $1$ or $1/3$ depending on the case.   

   We summarize the parameters of the different models explored in 
   Table~\ref{parameters}. We show the most relevant cases, in particular 
   those where X-ray emission is produced.\footnote{See \citet{ust16} for more 
   details about the formation of X-ray emitting shocks at magnetized protostellar jets.}
   
   \begin{table*}
      \caption[]{Summary of the initial physical parameters characterizing 
      the different models: 
      jet temperature at the lower boundary, $T_{\mathrm{j}}$; 
      jet density at the lower boundary (continuous component), $n_{\mathrm{j}}$;
      initial blob-to-jet particle number density ratio (pulsed component),
      $\chi_{\mathrm{b}} = n_{\mathrm{b}}/n_{\mathrm{j}}$;
      blob velocity at the lower boundary (pulsed component), $\varv_{\mathrm{b}}$; 
      initial blob-to-jet radius ratio (pulsed component), 
      $R_{\mathrm{b}}=r_{\mathrm{b}}/r_{\mathrm{j}}$; 
      a parameter, ROT, that indicates if there is rotation in the jet (``yes'') or not (``no'');
      and the mass loss rate calculated at the lower boundary, 
      $\dot M_{\mathrm{j}}$. 
      The initial temperature, density, and pressure of the ambient are 
      $T_{\mathrm{a}}=10\,$K, $n_{\mathrm{a}}=5000\,$cm$^{-3}$, and 
      $p_{\mathrm{a}}\approx 1.4\cdot 10^{-11}\,$dyne~cm$^{-2}$ respectively, in the LJ models; 
      and  $T_{\mathrm{a}}=100\,$K, $n_{\mathrm{a}}=100\,$cm$^{-3}$, and 
      $p_{\mathrm{a}}\approx 2.8\cdot 10^{-12}\,$dyne~cm$^{-2}$ in the HJ models. 
      The rest of the parameters are constant and equal for all the models: 
      ambient magnetic field, $B_{\mathrm{z}}=$ 4~mG; jet velocity at the lower boundary, 
      $\varv_{\mathrm{j}}=500\,$km~s$^{-1}$.}
      \label{parameters}
      \centering
      \begin{tabular}{p{3cm} ccccccc}
      \hline\hline
      Model  &  $T_{\mathrm{j}}$ (10$^{6}$ K)  &  
      $n_{\mathrm{j}}$ (10$^{4}$ cm$^{-3}$)  &  $\chi_{\mathrm{b}}$  &  
      $\varv_{\mathrm{b}}$ (km s$^{-1}$)  &  $R_{\mathrm{b}}$  &  ROT  &
      $\dot M_{\mathrm{j}}$ ($10^{-8} M_{\odot}$yr$^{-1}$)\\
      \hline
      \multicolumn{8}{c}{Light jet} \\
      \hline
      LJ1  &  $3$  &  $1$  &  $3$  &  300-700 {} \tablefootmark{c}  &  1  &  no  &  $1.3-5.4$ \\
      LJ2  &  $3$  &  $1$  &  $1.5$  &  300-700 {} \tablefootmark{c}  &  1  &  no  &  $1.3-2.7$ \\
      LJ3  &  $3$  &  $1$  &  $3$  &  500  &  1  &  no  &  $1.3-3.9$ \\
      LJ4  &  $3$  &  $1$  &  $3$  &  500  &  1/3  &  no  &  $1.3-1.6$ \\
      LJ5\_ref {} \tablefootmark{a}\tablefootmark{b}  &  $3$  &  $1$  &  $3$  &  500  &  1  &  yes  &  $1.4-4.0$ \\
      LJ6  &  $3$  &  $1$  &  $1.5$  &  500  &  1  &  no  &  $1.3-2.0$ \\
      LJ7  &  $3$  &  $1$  &  $1.5$  &  500  &  1/3  &  no  &  $1.3-1.4$ \\
      LJ8  &  $3$  &  $1$  &  $10$  &  500  &  1  &  no  &  $1.3-12.8$ \\
      LJ9  &  $3$  &  $2$  &  $1.5$  &  500  &  1  &  no  &  $2.6-3.9$ \\
      LJ10  &  $2$  &  $3$  &  $1.5$  &  500  &  1  &  no  &  $3.9-5.8$ \\
      \hline
      \multicolumn{8}{c}{Heavy jet} \\
      \hline
      HJ1  &  $1$  &  $1$  &  $3$  &  300-700 {} \tablefootmark{c}  &  1  &  no  &  $1.3-5.3$ \\
      HJ2 {} \tablefootmark{b}  &  $1$  &  $1$  &  $1.5$  &  300-700 {} \tablefootmark{c}  &  1  &  yes  &  $1.3-2.7$ \\
      HJ3\_ref {} \tablefootmark{a}\tablefootmark{b}  &  $1$  &  $1$  &  $1.5$  &  300-700 {} \tablefootmark{c}  &  
      1/3  &  yes  &  $1.3-1.5$ \\
      HJ4  &  $1$  &  $1$  &  $3$  &  500  &  1  &  no  &  $1.3-3.8$ \\
      HJ5  &  $1$  &  $1$  &  $3$  &  500  &  1/3  &  no  &  $1.3-1.6$ \\
      HJ6 {} \tablefootmark{b}  &  $1$  &  $1$  &  $3$  &  500  &  1  &  yes  &  $1.3-3.9$ \\
      HJ7  &  $1$  &  $1$  &  $1.5$  &  500  &  1  &  no  &  $1.3-1.9$ \\
      HJ8  &  $1$  &  $1$  &  $1.5$  &  500 &  1/3  &  no  &  $1.3-1.4$ \\
      HJ9  &  $1.5$  &  $1$  &  $1.5$  &  500  &  1  &  no  &  $1.3-1.9$ \\
      HJ10  &  $2$  &  $1$  &  $1.5$  &  500  &  1  &  no  &  $1.3-1.9$ \\
      HJ11  &  $1.5$  &  $1$  &  $3$  &  500  &  1  &  no  &  $1.3-3.8$ \\
      HJ12  &  $2$  &  $1$  &  $3$  &  500  &  1  &  no  &  $1.3-3.8$ \\
      \hline
      \end{tabular}
      \tablefoot{The top panel shows the models for the light jet, a jet less dense 
      than the ambient medium. The bottom panel displays the models for the heavy jet, 
      a jet denser than the ambient medium.\\
      \tablefoottext{a}{Reference case.}\\
      \tablefoottext{b}{Jet characterized by an angular velocity corresponding to 
      a maximum linear rotational velocity of 
      $\varv_{\varphi,\mathrm{max}} = 150$~km~s$^{-1}$ at the lower boundary.}
      \tablefoottext{c}{Random velocity with values between 300 and 700~km~s$^{-1}$.}
      }
   \end{table*}

   \subsection{Synthesis of X-ray emission}
   \label{sec:xray}
   
   We synthesize the 2D spatial distribution of count rate from the 
   simulations as follows. First we derive the 2D distributions of 
   temperature and density by integrating the MHD equations in the 
   whole spatial domain. Then we reconstruct the 3D spatial distributions 
   of these physical quantities by rotating the 2D slabs around the 
   symmetry axis $z$. For each cell of the 3D domain, we derive the 
   emission measure defined as 
   $EM = \int n_{\mathrm{e}} n_{\mathrm{H}} \mathrm{d}V$ 
   (where $n_{\mathrm{e}}$ and $n_{\mathrm{H}}$ are the electron and 
   hydrogen densities, respectively, and $V$ is the volume of emitting plasma).
   From the 3D spatial distributions of temperature and emission measure 
   reconstructed from the 2.5D simulations, we calculate the count rate in 
   the corresponding X-ray band filtering the emission through 
   \textit{Chandra}/ACIS instrumental response. For the light jet scenario we 
   consider an interestellar column density of 
   $N_{\mathrm{H}}=1.2\cdot 10^{22}$~cm$^{-2}$, best fit value determined 
   by \citet{bon11} for HH~154, while for the heavy jet scenario we assume a 
   value of $N_{\mathrm{H}}=1.5\cdot 10^{21}$~cm$^{-2}$, in good agreement 
   with values determined by \citet{gud05,gud08,gud11} for DG~Tau. 
   We also include Poisson fluctuations to mimic the photon count statistics. 
   The assumed exposure time is $t_{\mathrm{exp}}=100$~ks. 
   Then we derive the 2D distribution of the count rate by integrating along 
   the line of sight (assumed to be perpendicular to the jet axis). 
   Finally, in order to compare the images directly with the observations of 
   HH~154 and DG~Tau, we degrade the spatial resolution of the maps derived 
   from the simulations to 60~AU (\textit{Chandra} resolution at 140~pc).


\section{Results}

\subsection{Light jet: the case of HH~154}
\label{sec:LJ}

\begin{figure}
      \resizebox{\hsize}{!}{\includegraphics*{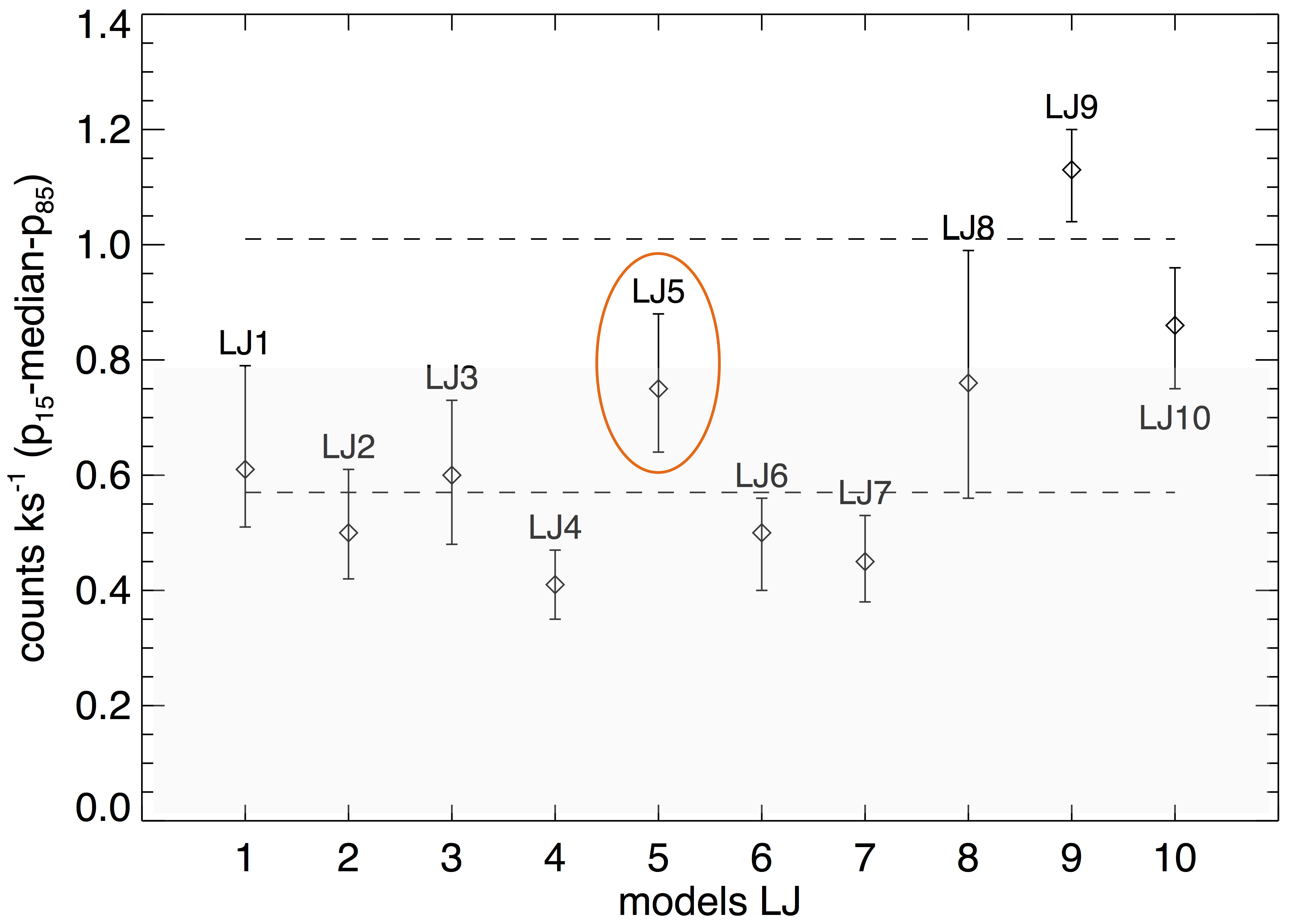}}
      \caption{X-ray count rate in the [0.3-4]~keV band of light jet models, 
      named in the horizontal axis as described in Table~\ref{parameters}. 
      In the vertical axis we plot the median count rate in every case 
      (represented with a diamond). 
      The lower and upper error bars indicate the 15\% and 85\% percentile 
      respectively. The dashed lines indicate the interval of the count rate 
      observed for HH~154. The orange circle indicates the reference case. 
      The shadowed part correspond to the scale of Fig.~\ref{sum_HJ} which
      summarizes the heavy jet models described in the next subsection.}
      \label{sum_LJ}
   \end{figure}

   Most of the models explored in the light jet scenario reproduce well 
   the case of HH~154. In Figure~\ref{sum_LJ} we summarize the results 
   of the count rate calculated for the different models described in 
   Table~\ref{parameters}.
   For every model we derive the count rate in the [0.3-4]~keV band during 
   the evolution (one value per year) as decribed in Section~\ref{sec:xray}, 
   and then we calculate the median and the 15\% and 85\% percentiles of 
   all the values. The median gives us a reference value for every model 
   while the percentiles are associated with the lower and upper variations 
   in every case. We indicate with dashed lines the interval containing the 
   X-ray count rate values (considering errors) derived from observations by 
   \citet{bon11}, namely 
   $0.76\pm0.10$, $0.65\pm0.08$ and $0.89\pm0.12$ counts ks$^{-1}$.
   This representation allows us to compare the different models between 
   them and with the values observed.
   The models that best fit with the HH~154 observations are LJ5, LJ8 and 
   LJ10 (see Figure~\ref{sum_LJ}). We assume LJ5 (marked with an 
   orange circle in Fig.~\ref{sum_LJ}) as our reference case and we 
   describe it in detail in the next section.
   
   \begin{figure*}
   \centering
      \includegraphics*[width=0.304\textwidth]{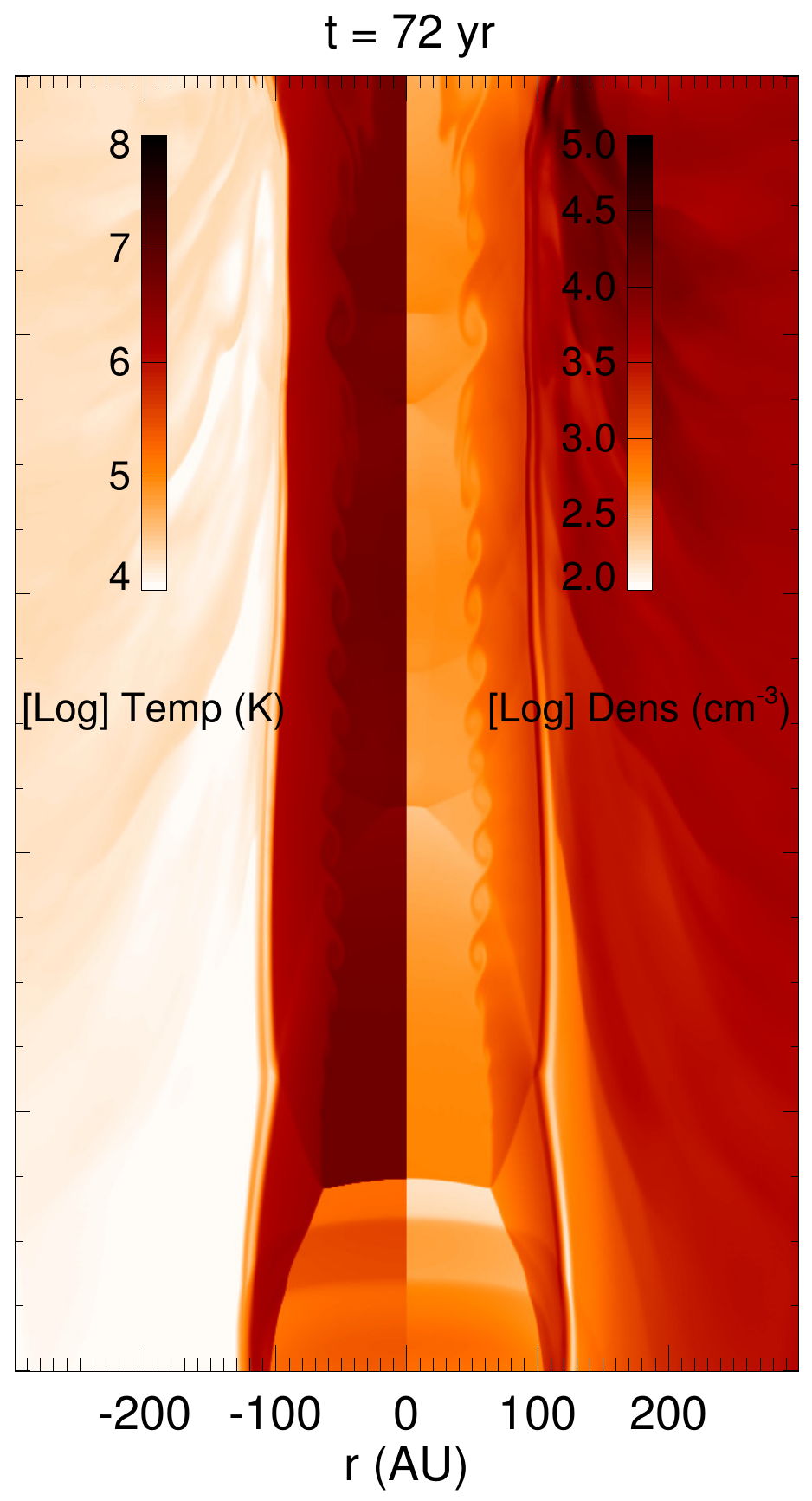}
      \includegraphics*[width=0.304\textwidth]{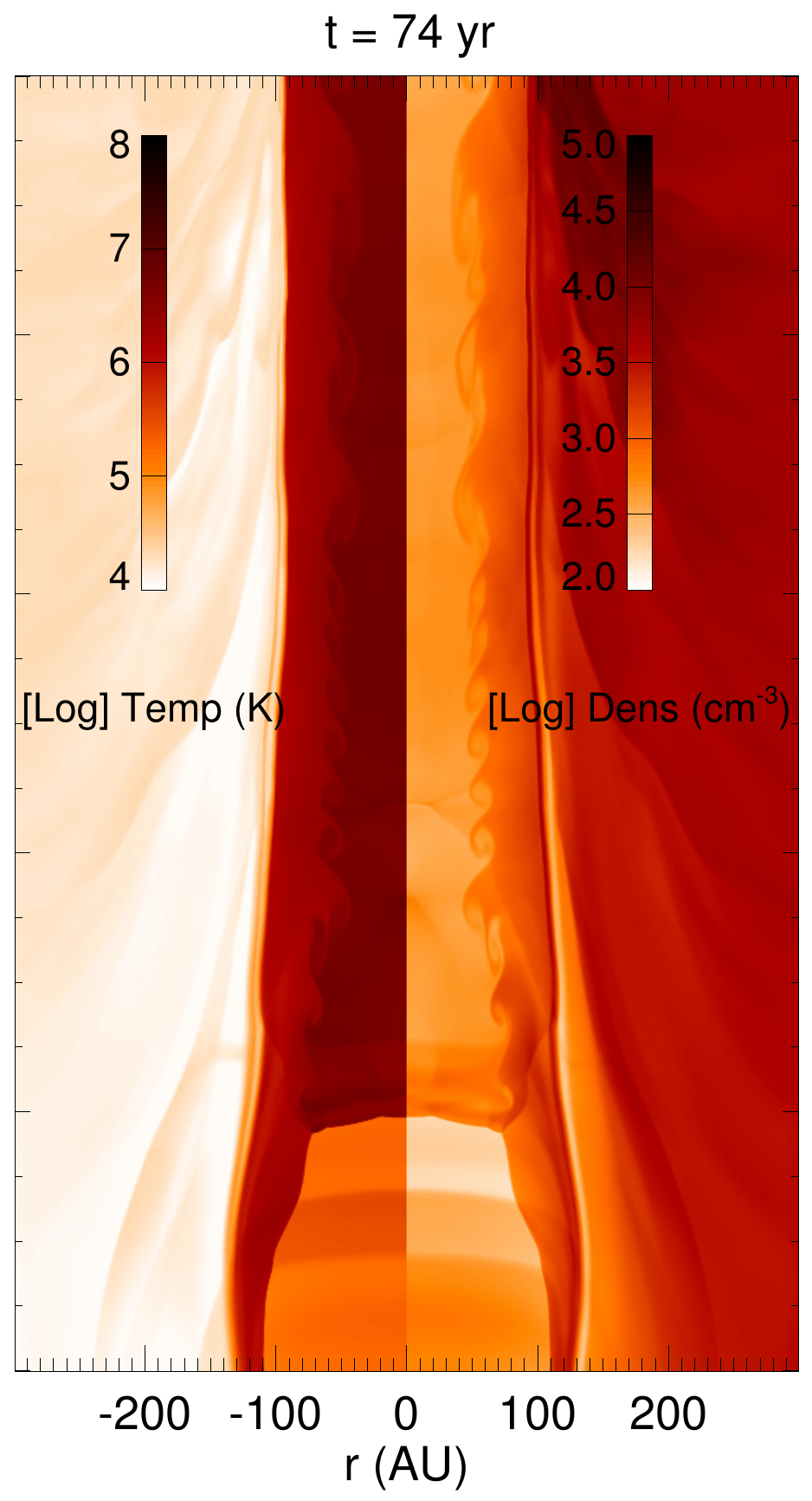}
      \includegraphics*[width=0.3805\textwidth]{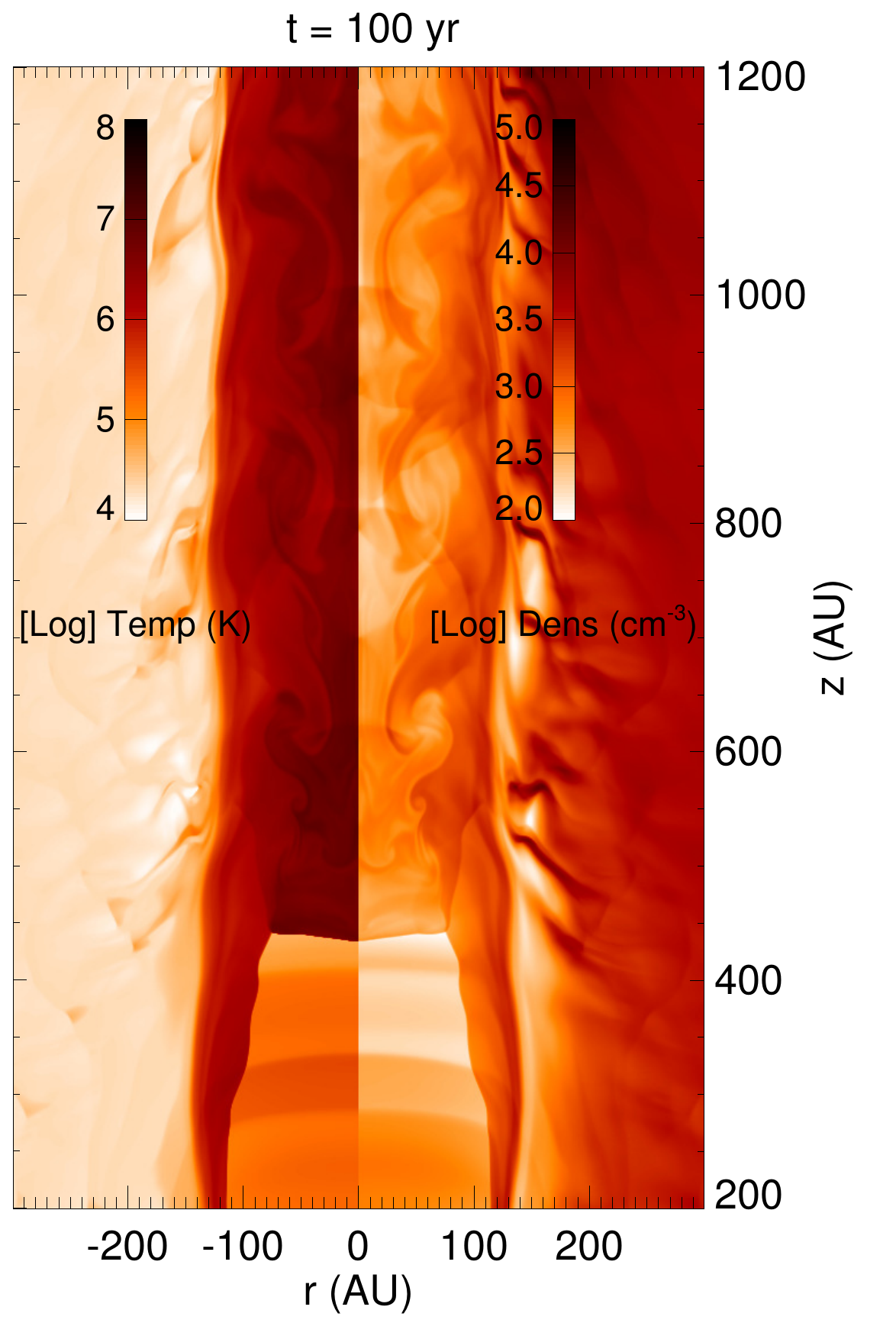}   
      \includegraphics*[width=0.304\textwidth]{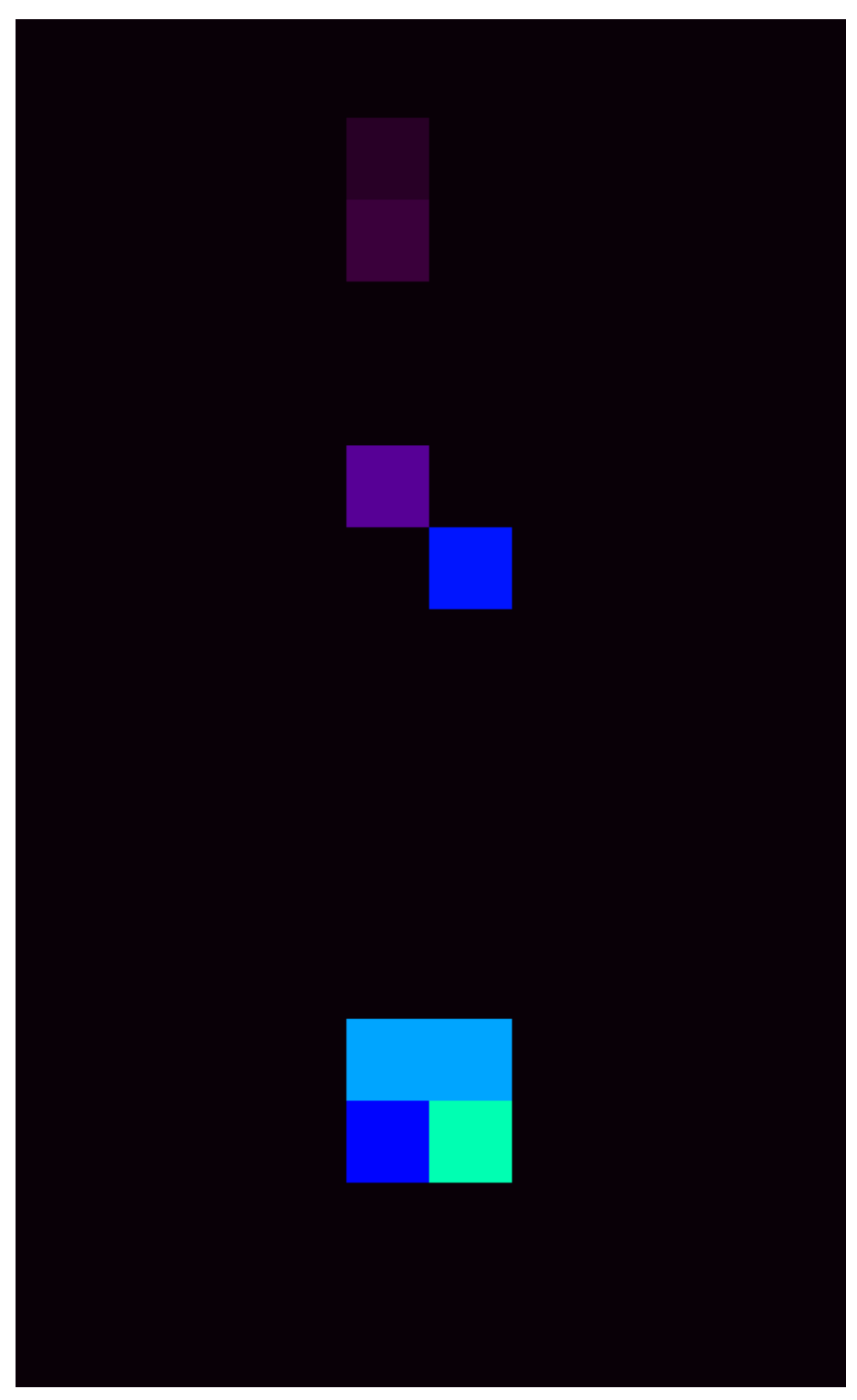}
      \includegraphics*[width=0.304\textwidth]{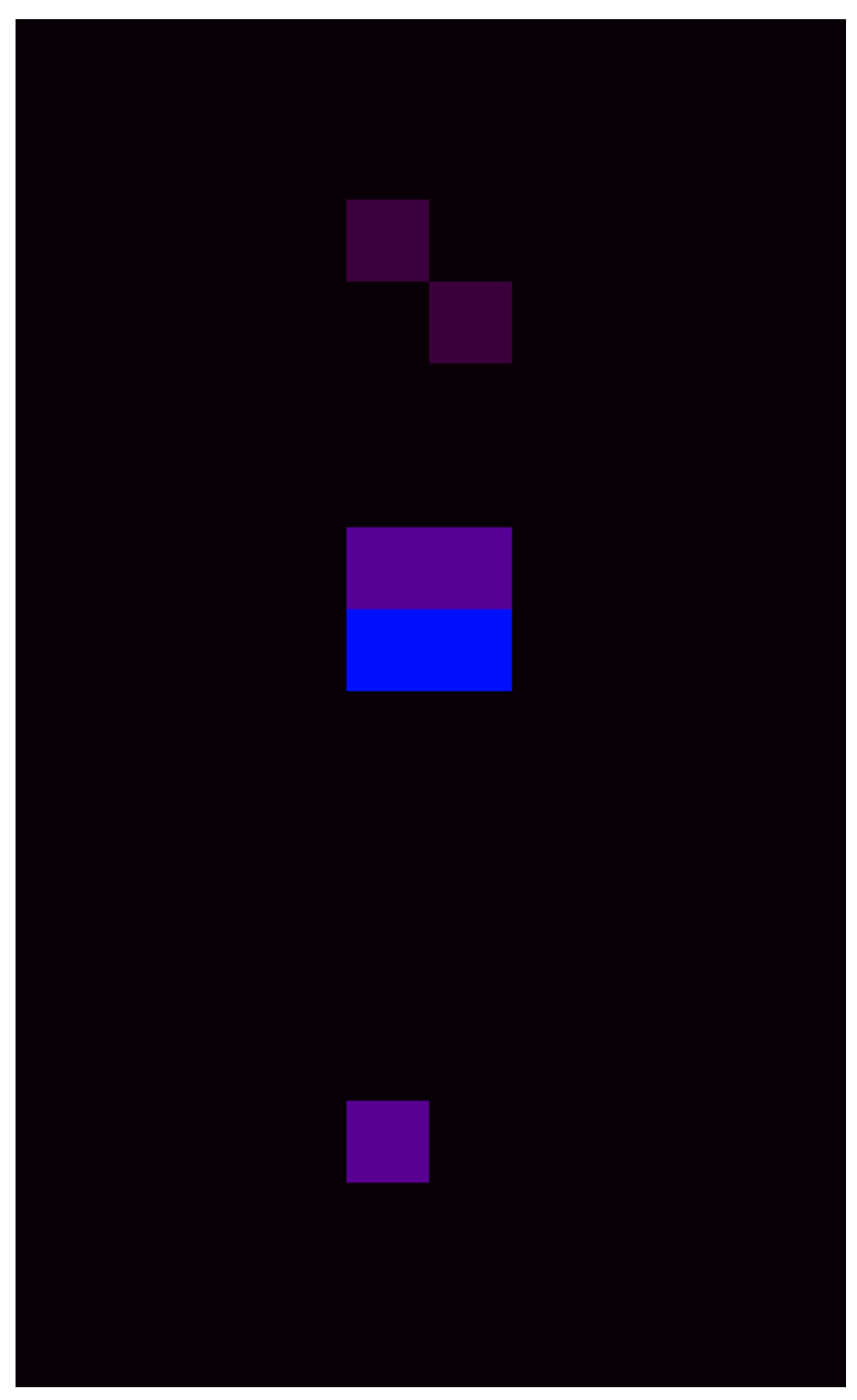}
      \includegraphics*[width=0.3805\textwidth]{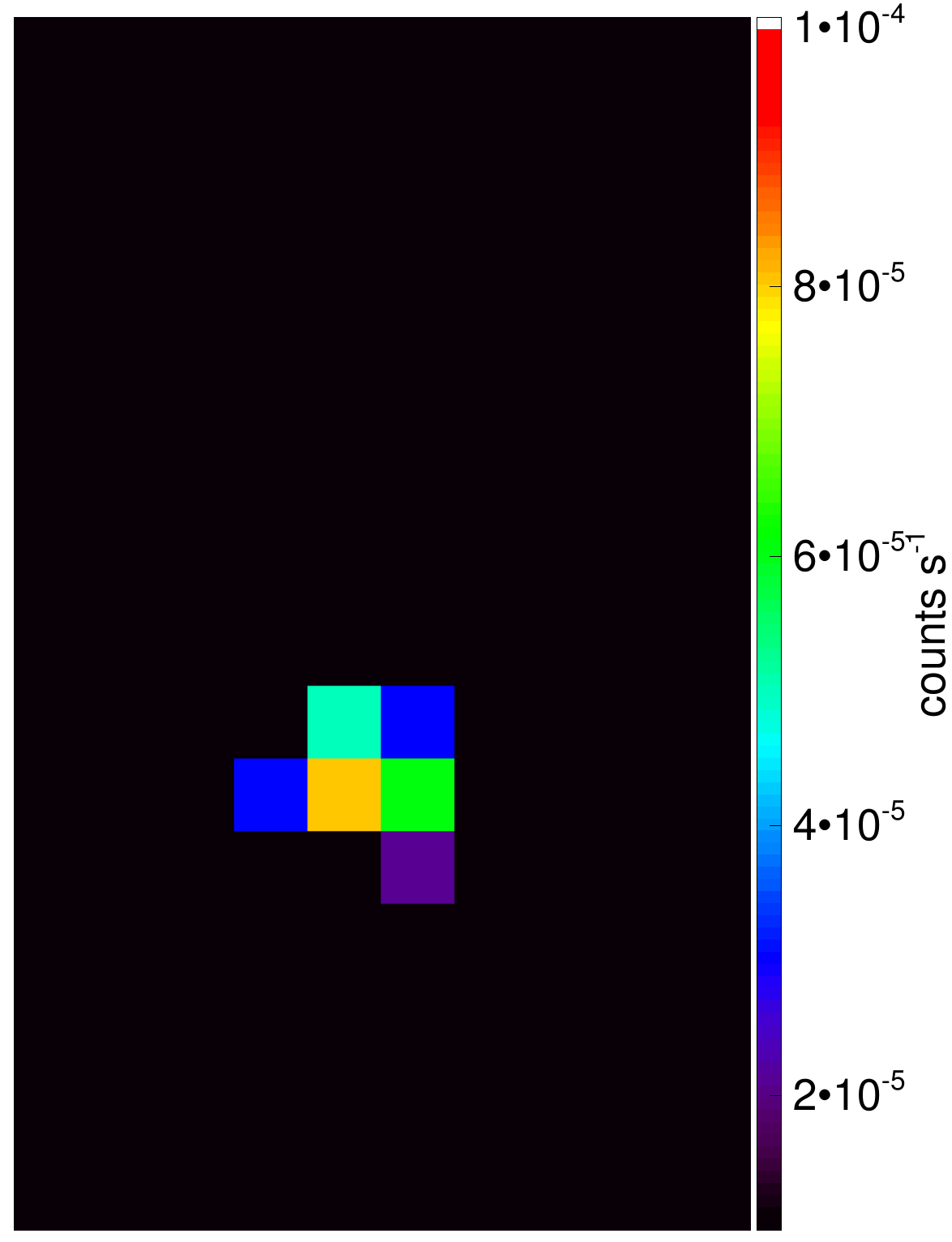}
      \caption{Model LJ5 at different evolution times: 
      $t \approx 72$~yr (left panels), $t \approx 74$~yr (middle panels), 
      and $t \approx 100$~yr (right panels). 
      Upper panels: Two-dimensional maps of temperature (left half-panels), 
      and density (right half-panels) distributions. 
      Lower panels: Maps of X-ray count rate in the [0.3-4]~keV band 
      with macropixel resolution of 0.5\arcsec.}
      \label{res_LJ}
   \end{figure*}
   
   \begin{figure*}
      \resizebox{\hsize}{!}{\includegraphics*{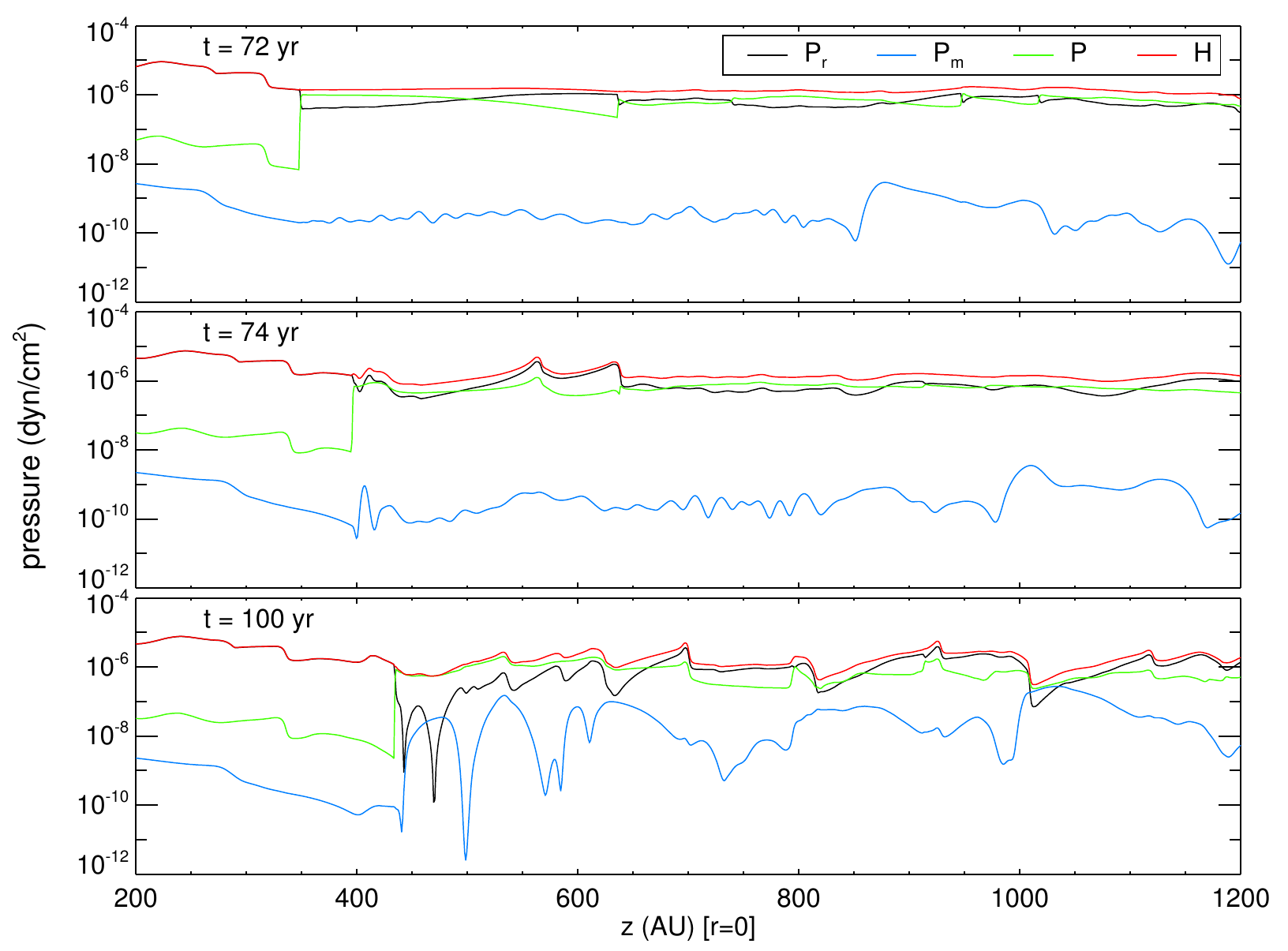}}
      \caption{Pressure profiles at $r = 0$ for the model LJ5, and 
      $t \approx 72$~yr (upper panel), $t \approx 74$~yr (middle panel), 
      and $t \approx 100$~yr (lower panel). We plot ram pressure in
      black, magnetic pressure in blue, thermal pressure in green 
      and dynamic pressure in red.}
      \label{plot_LJ}
   \end{figure*}
   
   \begin{figure}
      \resizebox{\hsize}{!}{\includegraphics*{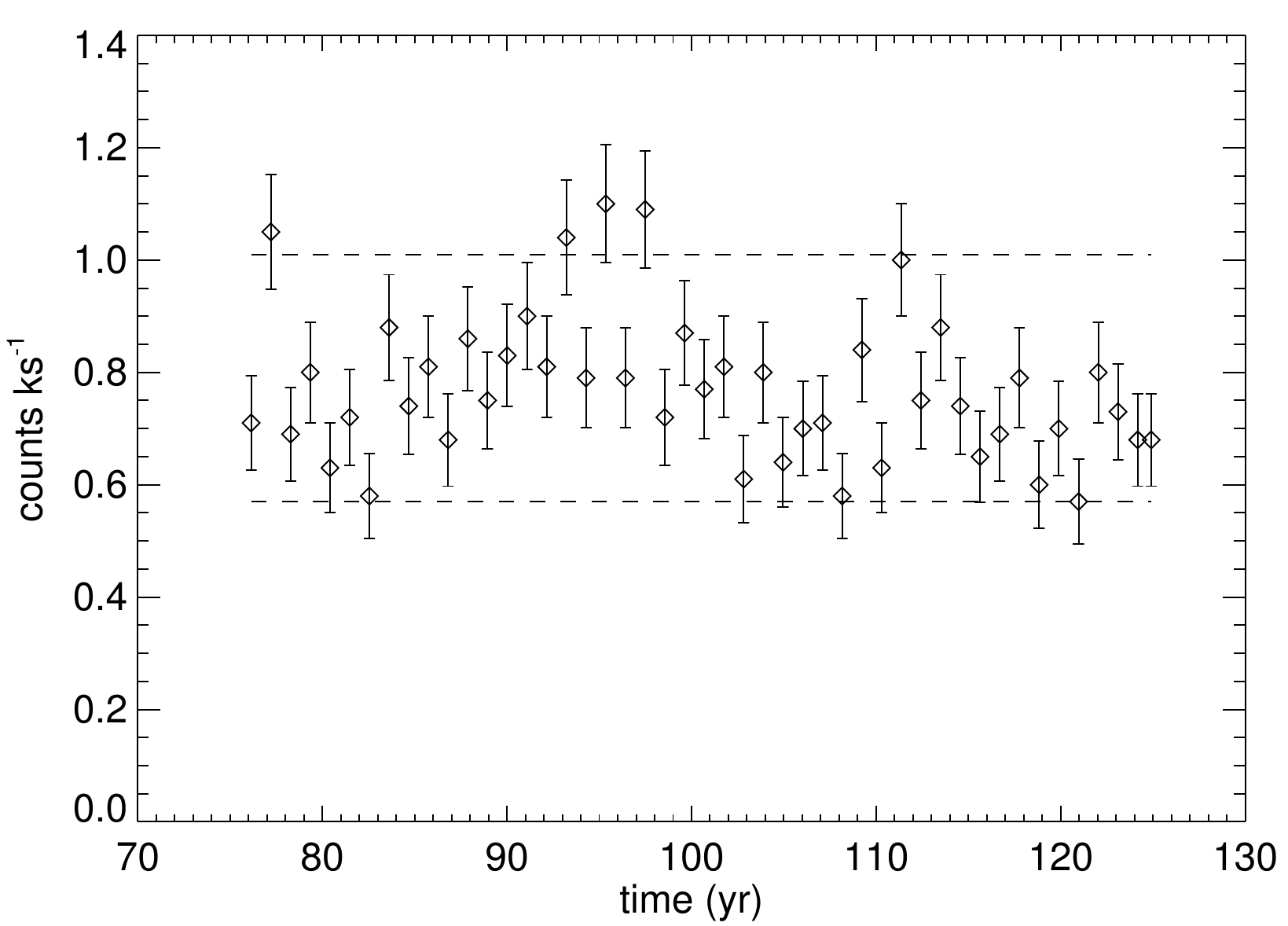}}
      \caption{X-ray count rate in the [0.3-4]~keV band with error bars 
      for the model LJ5. We plot one point every year. The dashed lines 
      indicate the interval of the count rate observed for HH~154.}
      \label{CRt_LJ}
   \end{figure}

   \begin{figure}
      \resizebox{\hsize}{!}{\includegraphics*{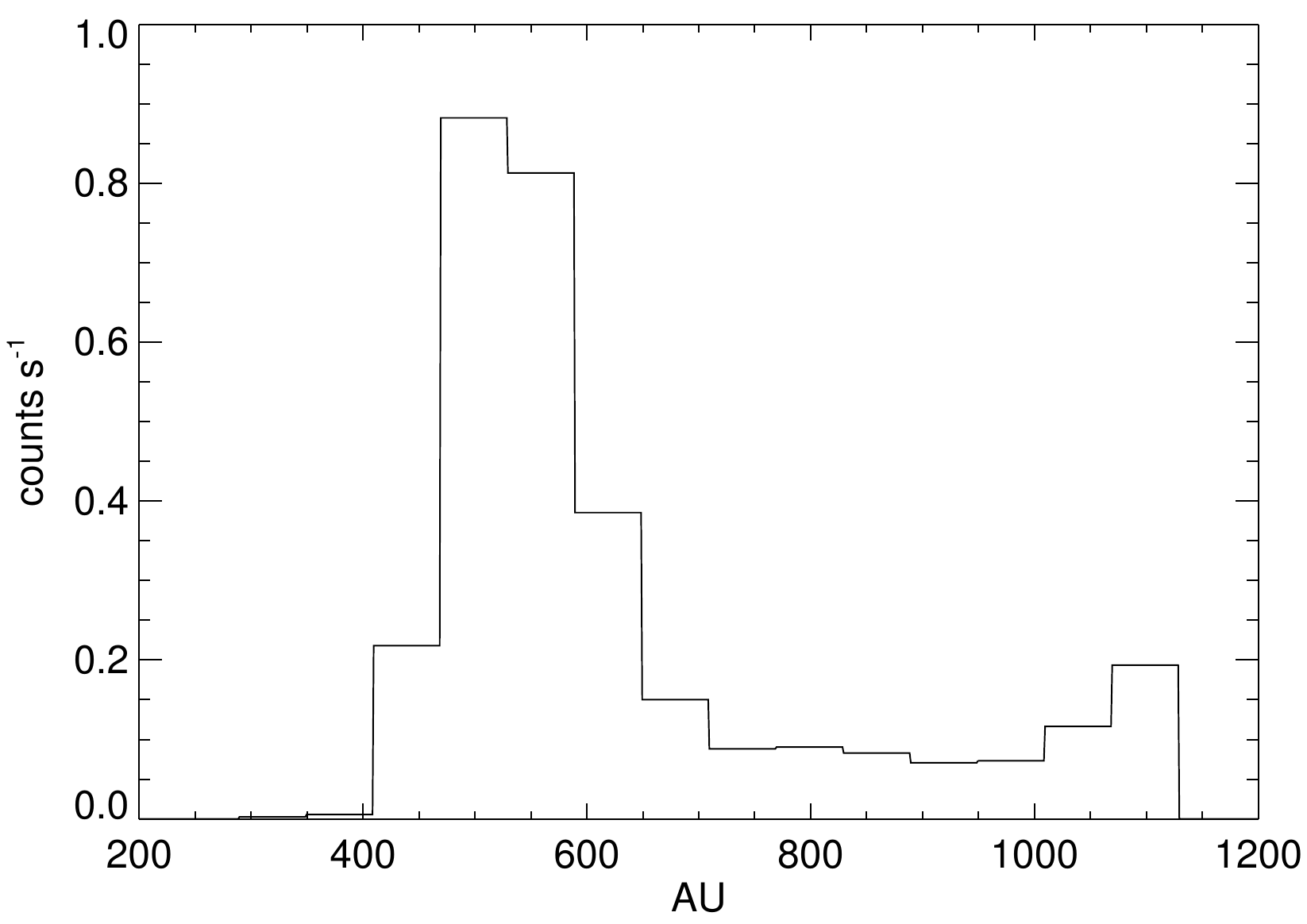}}
      \caption{X-ray count rate profile along the jet axis calculated integrating the 
      2D maps of the LJ5 model evolution in time and along the jet radius $r$.}
      \label{CRs2_LJ}
   \end{figure}

   \subsubsection{The reference case}
   
   In Figure~\ref{res_LJ} we report the 2D spatial distributions of temperature, 
   density, and count rate of the model LJ5 (see Table~\ref{parameters}) in 
   three different moments. The complete temporal evolution is available as 
   an online movie (Movie 1). The animation starts when the pulsed component
   is introduced, formed by a train of blobs with blob-to-jet particle number 
   density ratio $\chi_{\mathrm{b}} =3$ (see Table~\ref{parameters}), and 
   covers the evolution of the pulsed jet for approximately 50 years.
   The upper panels in Fig.~\ref{res_LJ} show 2D maps of temperature 
   (left half-panels) and density (right half-panels) both in logarithmic scale. 
   The lower panels show the 2D spatial distribution of X-ray count rate in 
   the [0.3-4]~keV band with resolution of 0.5\arcsec 
   (Chandra native resolution), derived from the simulations as described in 
   Section~\ref{sec:xray}.

   The left panels in Figure~\ref{res_LJ} show the stationary shock when 
   the first blob is arriving at $t \approx 72$~yr, for the model LJ5. 
   The shock forms when the flow expands and is collimated by the 
   ambient magnetic field heating the plasma to temperatures of a few million 
   degrees (see \citealt{ust16} for a detailed description).
   The plasma density and temperature reach respective maximum values 
   of $\sim 6\cdot10^{3}\,\mathrm{cm}^{-3}$ and 
   $\sim 7\cdot10^{6}\,\mathrm{K}$ at the shock. The shock temperature, 
   calculated as the density-weighted average temperature considering only 
   the cells with $T\geq 10^6$, is $\sim 3\cdot 10^6$~K. The pre-shock 
   density is $\sim 500\,\mathrm{cm}^{-3}$, one order of magnitude lower
   than the ambient medium density, namely $\sim 5000\,\mathrm{cm}^{-3}$.
   The count rate map in the [0.3-4]~keV band, calculated as described in 
   Section~\ref{sec:xray}, shows that the X-ray emission comes mainly from 
   the shock diamond (see lower left panel in Fig.~\ref{res_LJ}).
   The X-ray total shock luminosity, $L_{\mathrm{X}}$, derived in the 
   [0.3-4]~keV band is $\sim 5\cdot 10^{29}$~erg~s$^{-1}$.
   In Figure~\ref{plot_LJ} (upper panel) we plot the pressure profiles along 
   the jet axis at $t \approx 72$~yr. Close to the jet axis the model evolution 
   is dominated by the jet plasma pressure ($P$, represented in green) over 
   the magnetic pressure ($P_{\mathrm{m}}$, represented in blue), where 
   the plasma $\beta$ (defined as the ratio of the plasma pressure to the 
   magnetic pressure) is higher than 1 ($\beta>1$). In black, we plot the ram 
   pressure defined as $P_{\mathrm{r}} = \rho \cdot u^2$, where $\rho$ 
   and $u$ are the density and velocity respectively. Finally, we represent in 
   red the dynamic pressure, $H = P+P_{\mathrm{r}}$, almost constant 
   along the profile due to the stability and quasi-stationarity of the model.    
   The central panels in Figure~\ref{res_LJ} show the jet at $t \approx 74$~yr, 
   when the first blob just passed through the shock and the second one 
   is arriving. The shocked plasma density and temperature mantain mostly 
   similar values as before: maximum values of 
   $\sim 6\cdot10^{3}\,\mathrm{cm}^{-3}$ and $\sim 7\cdot10^{6}\,\mathrm{K}$,
   and density-weighted average temperature of $\sim 3\cdot 10^6$~K. 
   The X-ray source is perturbed and moved by the blob (see lower middle 
   panel in Fig.~\ref{res_LJ}).
   The X-ray total shock luminosity, $L_{\mathrm{X}}$, derived in the 
   [0.3-4]~keV band is $\sim6\cdot 10^{29}$~erg~s$^{-1}$.
   In Figure~\ref{plot_LJ} (middle panel) we observe the perturbation effect in 
   the pressure profiles along the jet axis at $t \approx 74$~yr, when the blob 
   has just passed, affecting slightly the shock stability.  
   The right panels in Figure~\ref{res_LJ} show the jet at $t \approx 100$~yr, 
   after a train of blobs passed through the shock. The shocked plasma 
   maximum values for density and temperature are 
   $\sim 3\cdot10^{3}\,\mathrm{cm}^{-3}$ and $\sim 1\cdot10^{7}\,\mathrm{K}$
   respectively, and the density-weighted average temperature is 
   $\sim 3\cdot 10^6$~K. The X-ray emission in the [0.3-4]~keV band is enhanced 
   by the perturbations and the source is located at the base of the jet (see lower 
   right panel in Fig.~\ref{res_LJ}). The X-ray total shock luminosity, $L_{\mathrm{X}}$, 
   derived in the [0.3-4]~keV band is $\sim7\cdot 10^{29}$~erg~s$^{-1}$. 
   In Figure~\ref{plot_LJ} (lower panel) we observe the pressure profiles along 
   the jet axis at $t \approx 100$~yr, completely perturbed by the train of blobs.

   \subsubsection{Variability}
   
   In order to investigate the stationarity of the different pulsed jet 
   models, we study the variations of the physical quantities (shock temperature, 
   density of the X-ray emitting component, X-ray luminosity, etc.) and of the 
   spatial distribution of the count rate during the evolution of the model. 
   We derive the total count rate in the [0.3-4]~keV band, calculated as 
   described in Sect.~\ref{sec:xray}, and integrated in all the domain to 
   obtain the total value. In Figure~\ref{CRt_LJ} we show
   one value of count rate with Poisson error bars per year, omitting the 
   first five values corresponding to the intial transient of the pulsed jet.
   The dashed lines correpond to the interval [0.57,1.01], which contains 
   the count rate values derived from observations by \citet{bon11}.
   We note that the values derived from the model are compatible with the 
   observations during all the evolution.
   The temporal evolution of the spatial distribution of the count rate for the 
   model LJ5, as would be seen by the \textit{Chandra}/ACIS instrument, 
   is reported in the right panel of the first movie (Movie 1). We observe two 
   different components emitting in X-rays during all the animation: one 
   quasi-stationary at the base of the jet, and another fainter in the direction 
   of propagation of the jet. In order to understand the trend of the X-ray 
   emission we integrate in time and along the jet radius, omitting the 
   frames corresponding to the initial transient. In this way, we derive the 
   count rate profile along the jet axis (see Fig.~\ref{CRs2_LJ}). 
   We find that most of the X-ray emission starkly comes from the source 
   close to the base of the jet. We also discern a fainter X-ray source 
   further away, as observed in HH~154 by \citet{bon11}.

   \subsubsection{Comparison with the other models}
   
   We compare different models for the light jet scenario in order to 
   investigate the effect of perturbations in the stationarity of the shock.
   The explored parameters are listed in Table~\ref{parameters}, namely jet 
   temperature and density, blob density, velocity and radius, the rotation,
   and the mass loss rate. The jet temperature and density, and the possible
   rotation, affect both continuous and pulsed components, and determine
   the physical parameters of the shock \citep[see][]{ust16}. They are 
   selected according to the observations of HH~154 \citep{fav02,fav06,bon11}. 
   The parameters of the blob (density, velocity and radius) are those that 
   determine the strength of the perturbation. The mass loss rate is calculated
   as explained in Section~\ref{sec:description} and give information about 
   the intensity of the perturbations introduced through the amplitude of its 
   variation. The values obtained for all the models are of the order of 
   $\sim 10^{-8}\,M_{\odot}$yr$^{-1}$, in good agreement with typical outflow 
   rates found in pre-main sequence stars \citep{cab07,pod11}.
   In our reference model (LJ5) the jet initial temperature and density are 
   $T_{\mathrm{j}}=3\cdot 10^{6}$~K and $n_{\mathrm{j}}=10^{4}$~cm$^{-3}$
   respectively. In cases with higher values of jet temperature and density, 
   namely LJ9 and LJ10, the shock is stronger and the count rate is higher 
   (see Fig.~\ref{sum_LJ}). In the model LJ3, with same parameters 
   as LJ5 but without rotation, the count rate is lower because the shock is 
   weaker as it was already predicted by \citet{ust16}. We do not observe 
   significant effect in the stability of the shock due to the rotation. 
   The parameters that directly affect the stationarity of the emitting shock 
   are those that define the blobs perturbing it. The parameters defining the 
   blob density, velocity and radius in the models LJ5 (reference case) and 
   LJ3 are $\chi_{\mathrm{b}}=3$, $\varv_{\mathrm{b}}=500$~km~s$^{-1}$ and 
   $R_{\mathrm{b}}=1$ respectively (see Table~\ref{parameters}). 
   When the blob density is lower (e.g. in LJ6, $\chi_{\mathrm{b}}=1.5$), 
   the median count rate and the perturbations are lower, whereas for a higher 
   blob density (e.g. in LJ8, $\chi_{\mathrm{b}}=10$), they are higher 
   comparing with LJ3.
   When the radius is lower (e.g. LJ4 with respect to LJ3, and LJ7 with 
   respect to LJ6) the median count rate and the perturbations are lower. 
   This effect is more evident in LJ4 because the perturbation is stronger.
   Finally, when we compare models with random velocity with those with 
   constant velocity (e.g. LJ1 with respect to LJ3, and LJ2 with respect 
   to LJ6), we do not observe a significant change.
   
   In summary, we can afirm that the perturbations are compatible with the 
   available observations of HH 154. The jet forms a quasi-stationary X-ray 
   emitting shock at the base of the jet and the perturbations arriving from 
   the protostellar source as a train of blobs contribute to the emission 
   enhancing the total count rate.
   The variations registered in the count rate are comparable with those 
   observed for perturbations in the following ranges: density increase of 
   a maximum of one order of magnitude, velocity fluctuation of ~50\% and 
   radius with values from 1/3-to-1 jet radius are compatible. The maximum 
   change in the mass loss rate derived is approximately one order of 
   magnitude in the model LJ8, for which the total count rate calculated 
   variations are at the limit of the values observed by \cite{bon11}.

\subsection{Heavy jet: the case of the jet associated to DG~Tau}
\label{sec:HJ}   

\begin{figure}
      \resizebox{\hsize}{!}{\includegraphics*{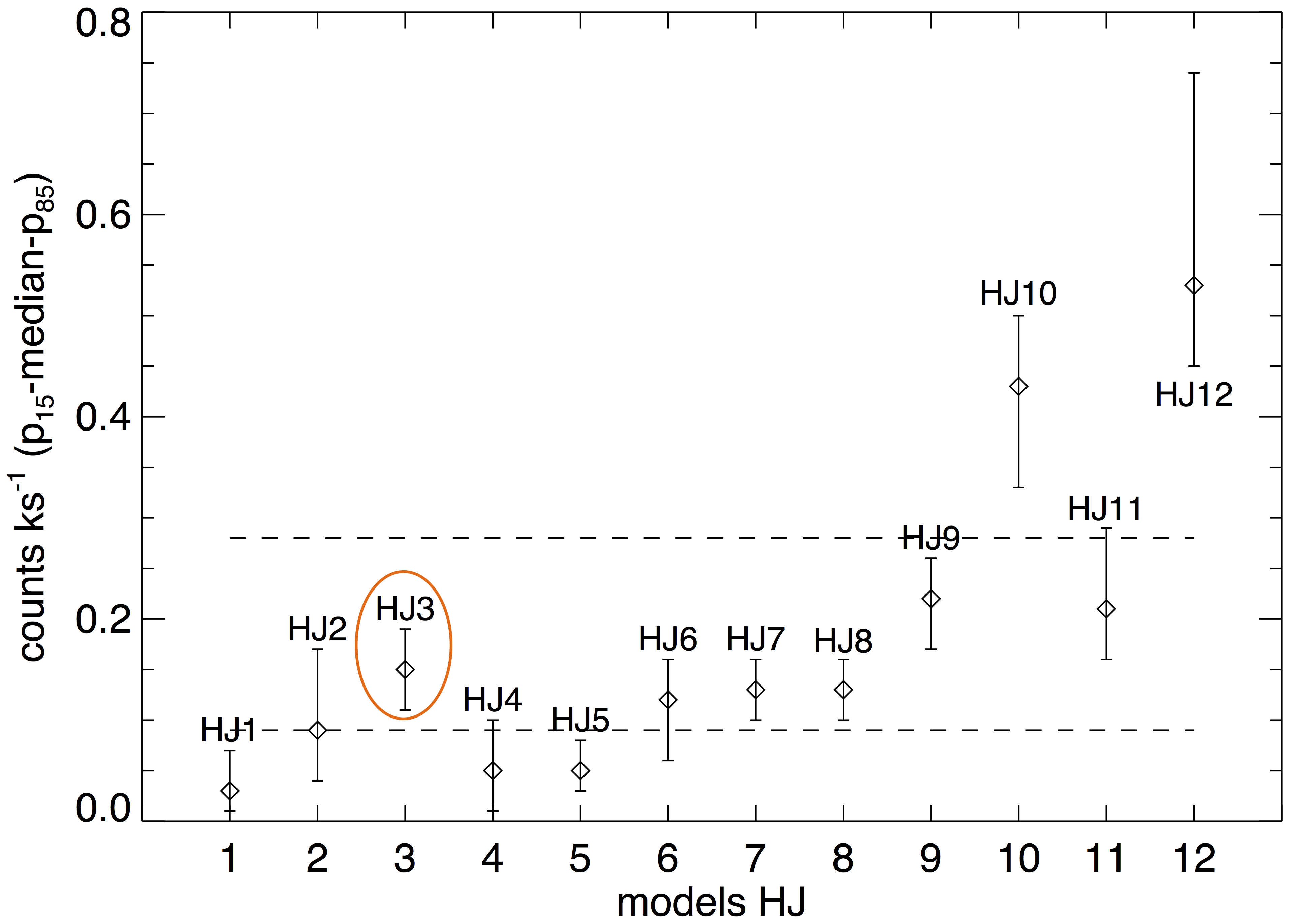}}
      \caption{X-ray count rate in the [0.5-1]~keV band of heavy jet models, 
      named in the horizontal axis as described in Table~\ref{parameters}. 
      In the vertical axis we plot the median count rate in every case 
      (represented with a diamond). 
      The lower and upper error bars represent the 15\% and 85\% percentile 
      respectively. The dashed lines indicate the interval of the count rate 
      observed for DG~Tau. The orange circle indicates the reference case.}
      \label{sum_HJ}
   \end{figure}

   Most of the models explored in the heavy jet scenario reproduce well the 
   case of DG Tau. In Figure~\ref{sum_HJ} we summarize the results for 
   the count rate for the different models descibed in Table~\ref{parameters}.
   For every model we derive the X-ray count rate in the [0.5-1]~keV band 
   during the evolution (one value per year) as decribed in 
   Section~\ref{sec:xray}, and then we calculate the median and the 15\% 
   and 85\% percentiles of all the values in each case. The dashed lines 
   indicate the interval containing the X-ray count rate values 
   (considering errors) derived from observations 
   by \citet{gud05,gud08,gud11}, namely 
   $0.20\pm0.08$, $0.18\pm0.06$ and $0.11\pm0.02$ counts ks$^{-1}$.
   As in the light jet scenario, this representation allows us to compare the 
   different models between them and with the values observed.
   The models that best fit with the DG~Tau observations are HJ3, HJ7, HJ8 
   and HJ9 (see Figure~\ref{sum_HJ}). We assume HJ3 (marked with 
   an orange circle in Fig.~\ref{sum_HJ}) as our reference case and we 
   describe it in detail in the next section.
   
   \begin{figure*}
   \centering
      \includegraphics*[width=0.304\textwidth]{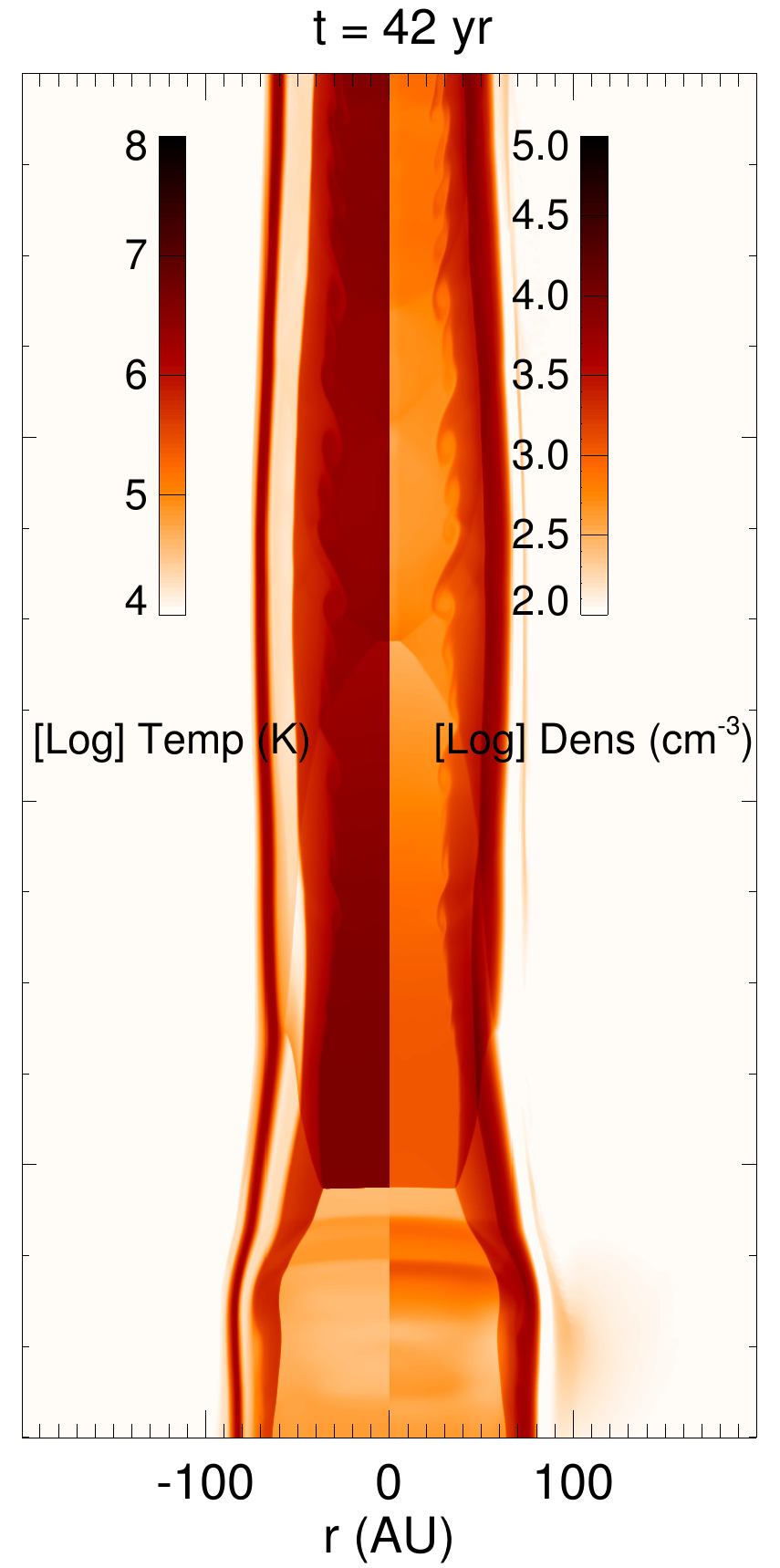}
      \includegraphics*[width=0.304\textwidth]{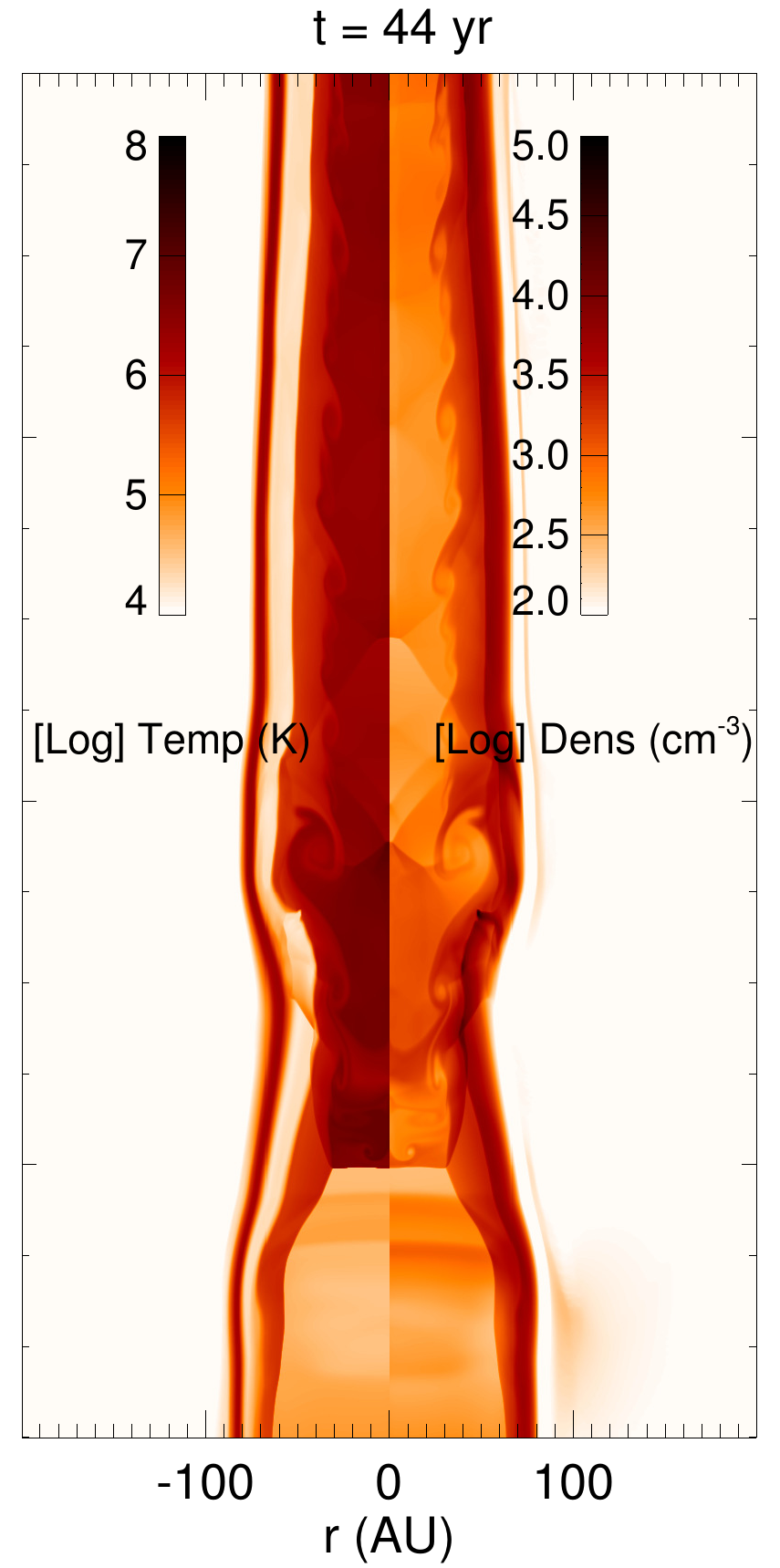}
      \includegraphics*[width=0.3805\textwidth]{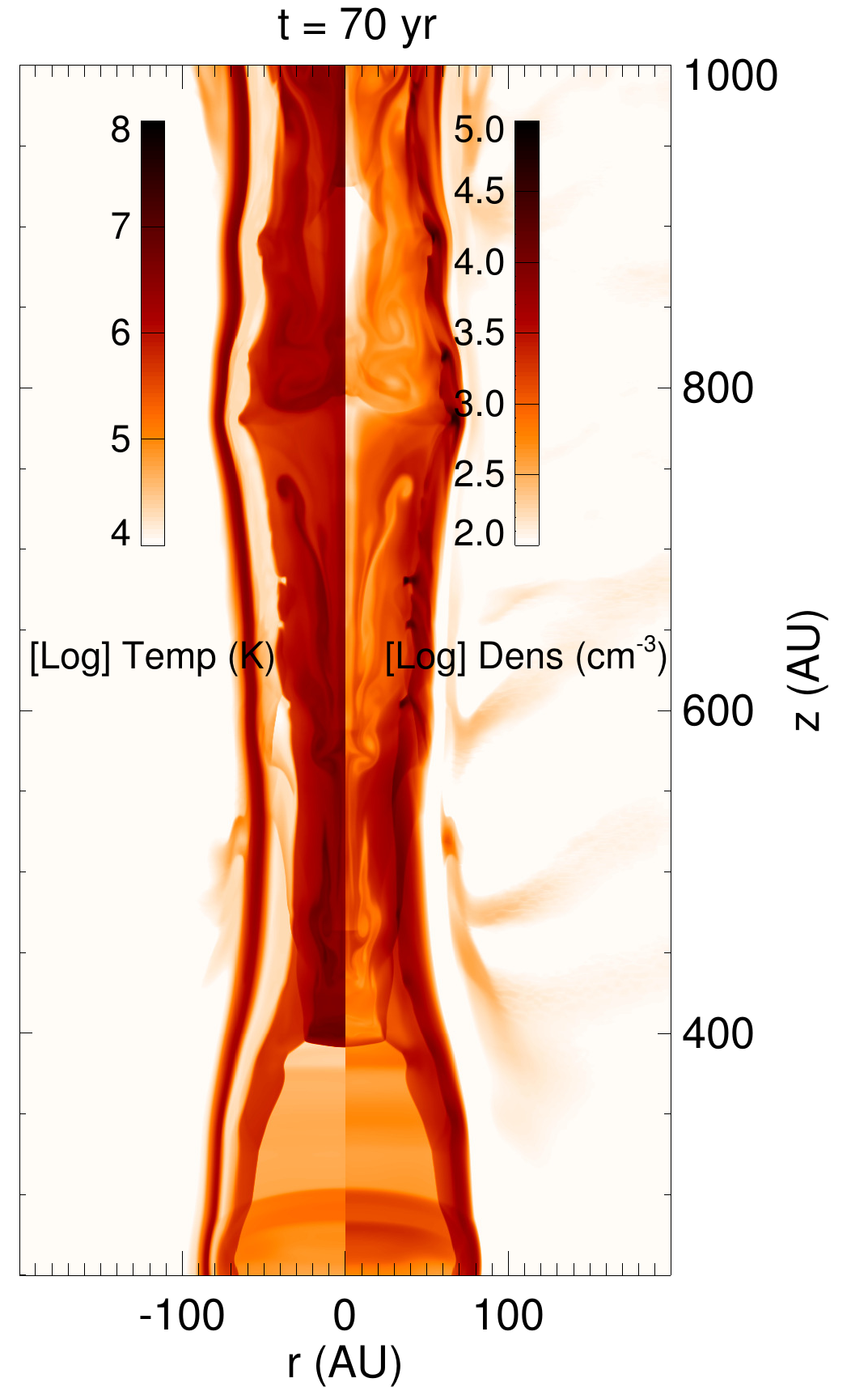}   
      \includegraphics*[width=0.304\textwidth]{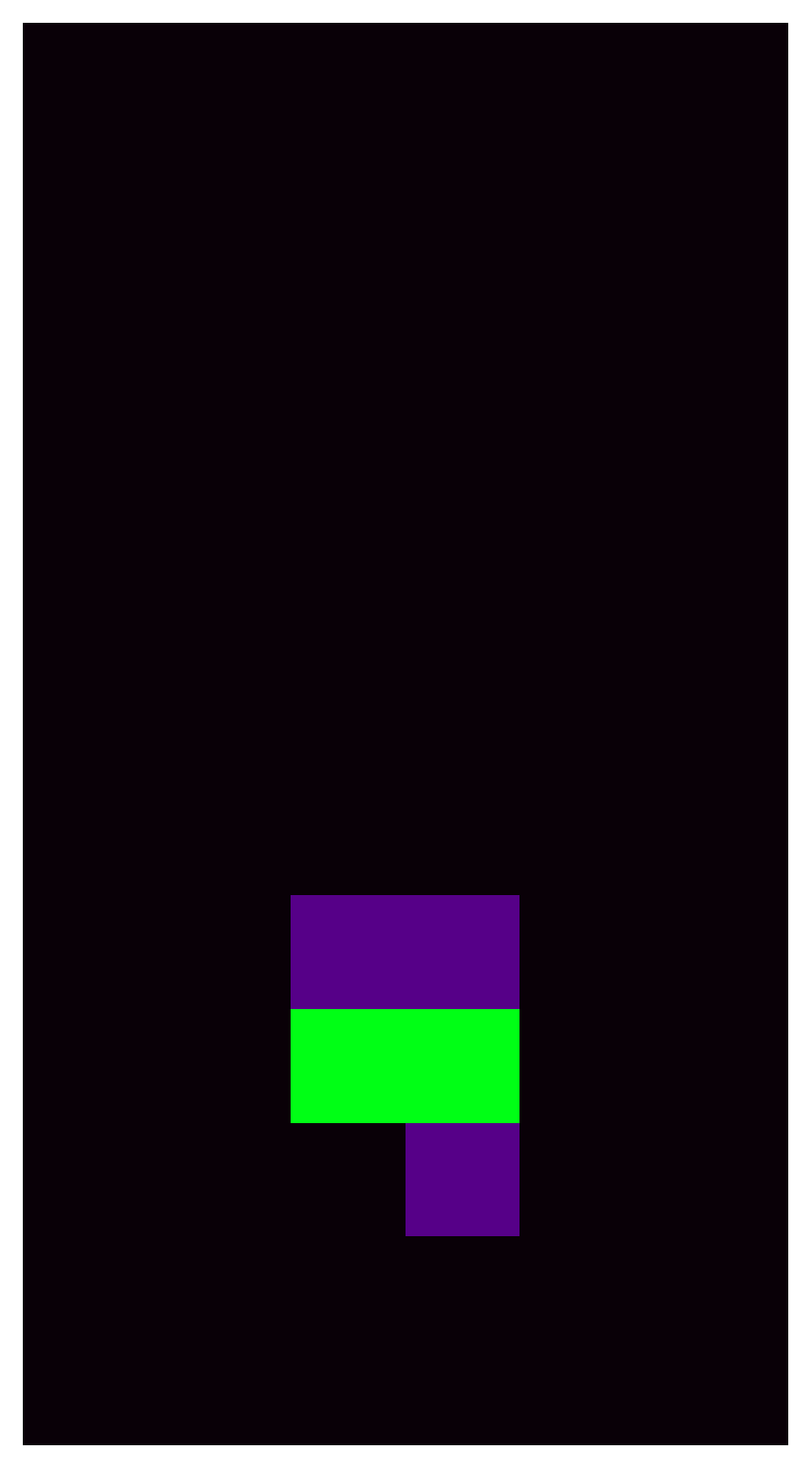}
      \includegraphics*[width=0.304\textwidth]{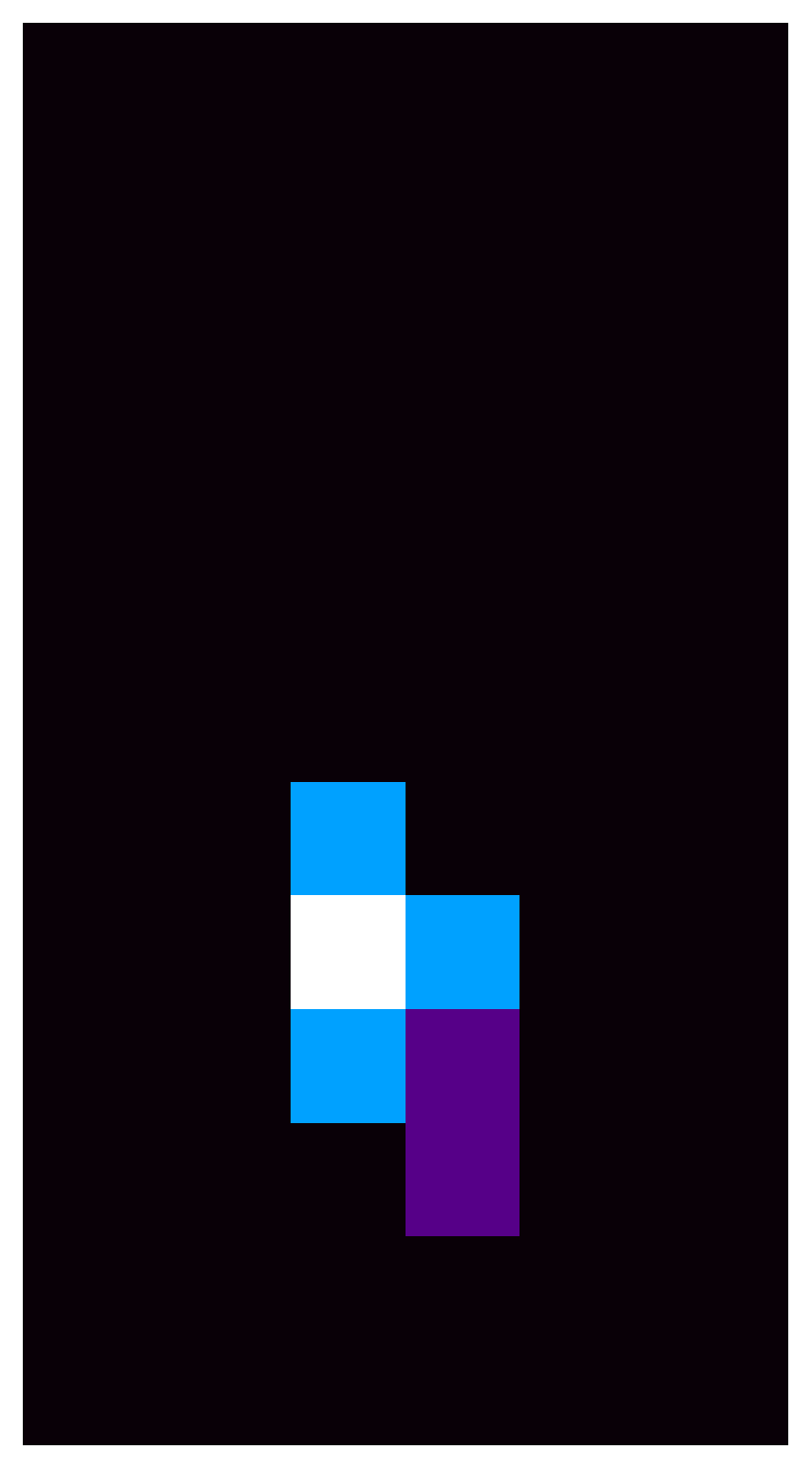}
      \includegraphics*[width=0.3805\textwidth]{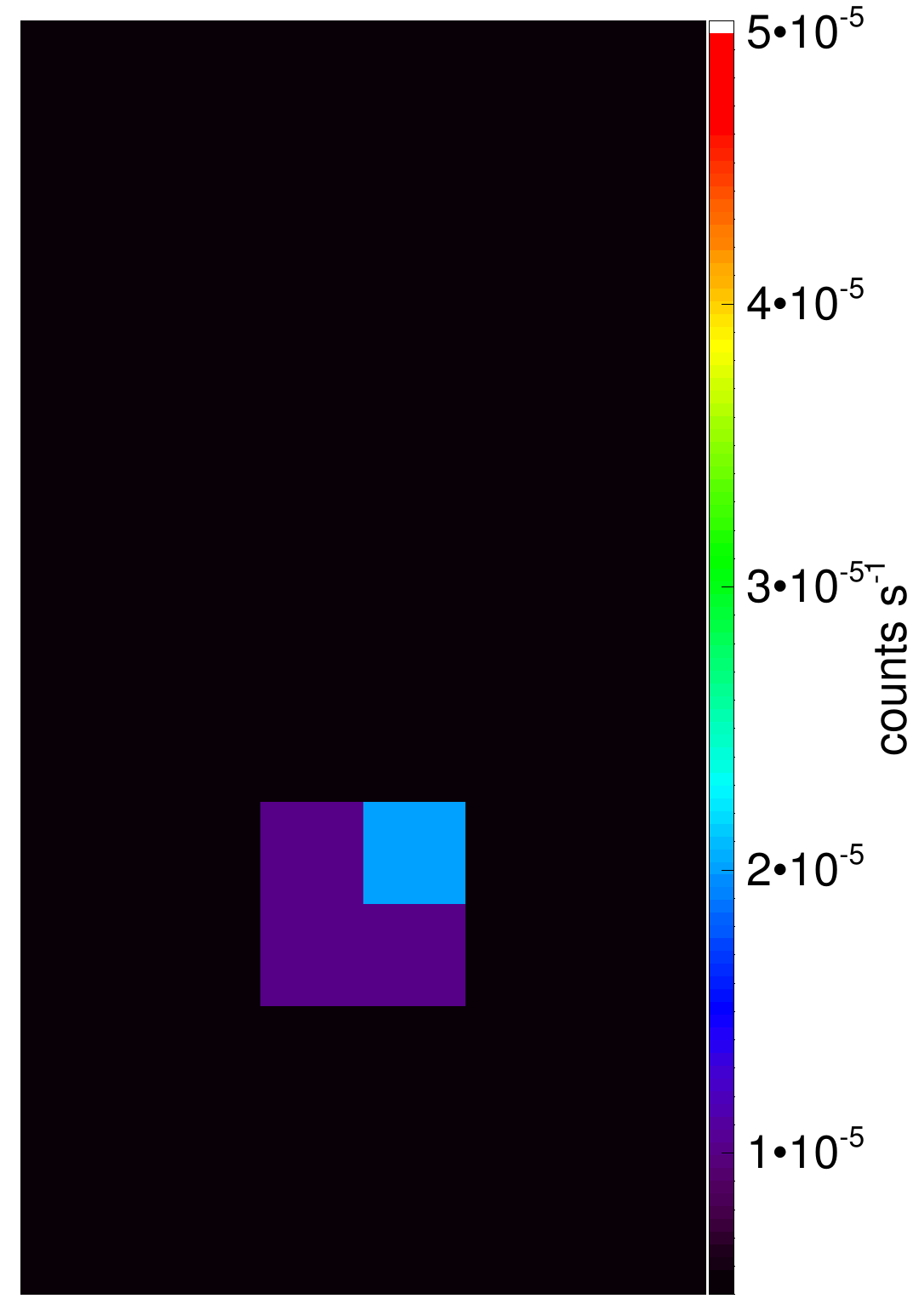}
      \caption{Model HJ3 at different evolution times: 
      $t \approx 42$~yr (left panels), $t \approx 44$~yr (middle panels), 
      and $t \approx 70$~yr (right panels). 
      Upper panels: Two-dimensional maps of temperature (left half-panels), 
      and density (right half-panels) distributions. 
      Lower panels: Maps of X-ray count rate in the [0.5-1]~keV band 
      with macropixel resolution of 0.5\arcsec.}
      \label{res_HJ}
   \end{figure*}
   
   \begin{figure*}
      \resizebox{\hsize}{!}{\includegraphics*{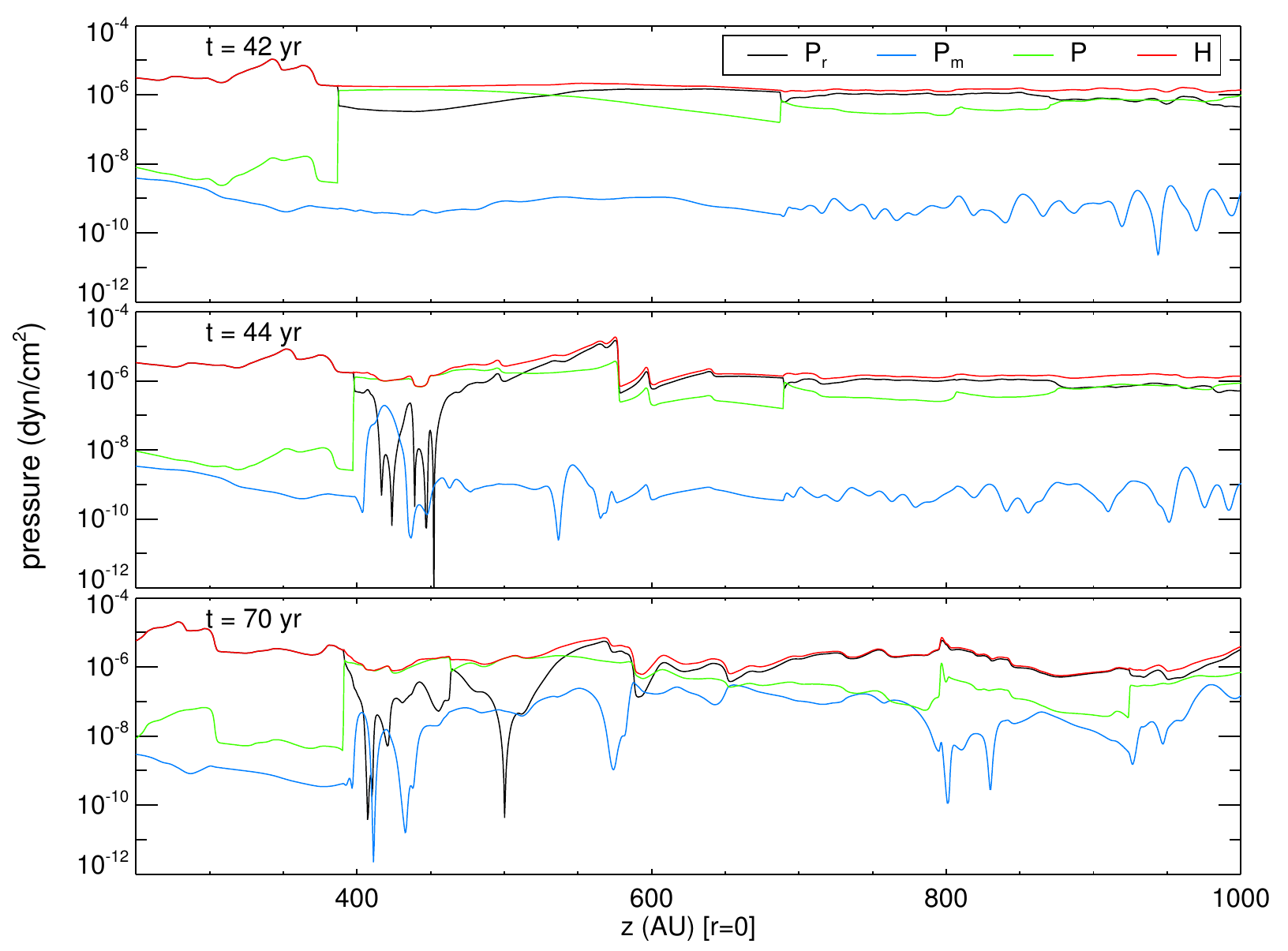}}
      \caption{Pressure profiles at $r = 0$ for the model HJ3, and 
      $t \approx 42$\,yr (upper panel), $t \approx 44$\,yr (middle panel), 
      and $t \approx 70$\,yr (lower panel). We plot ram pressure in
      black, magnetic pressure in blue, thermal pressure in green 
      and dynamic pressure in red.}
      \label{plot_HJ}
   \end{figure*}
   
   \begin{figure}
      \resizebox{\hsize}{!}{\includegraphics*{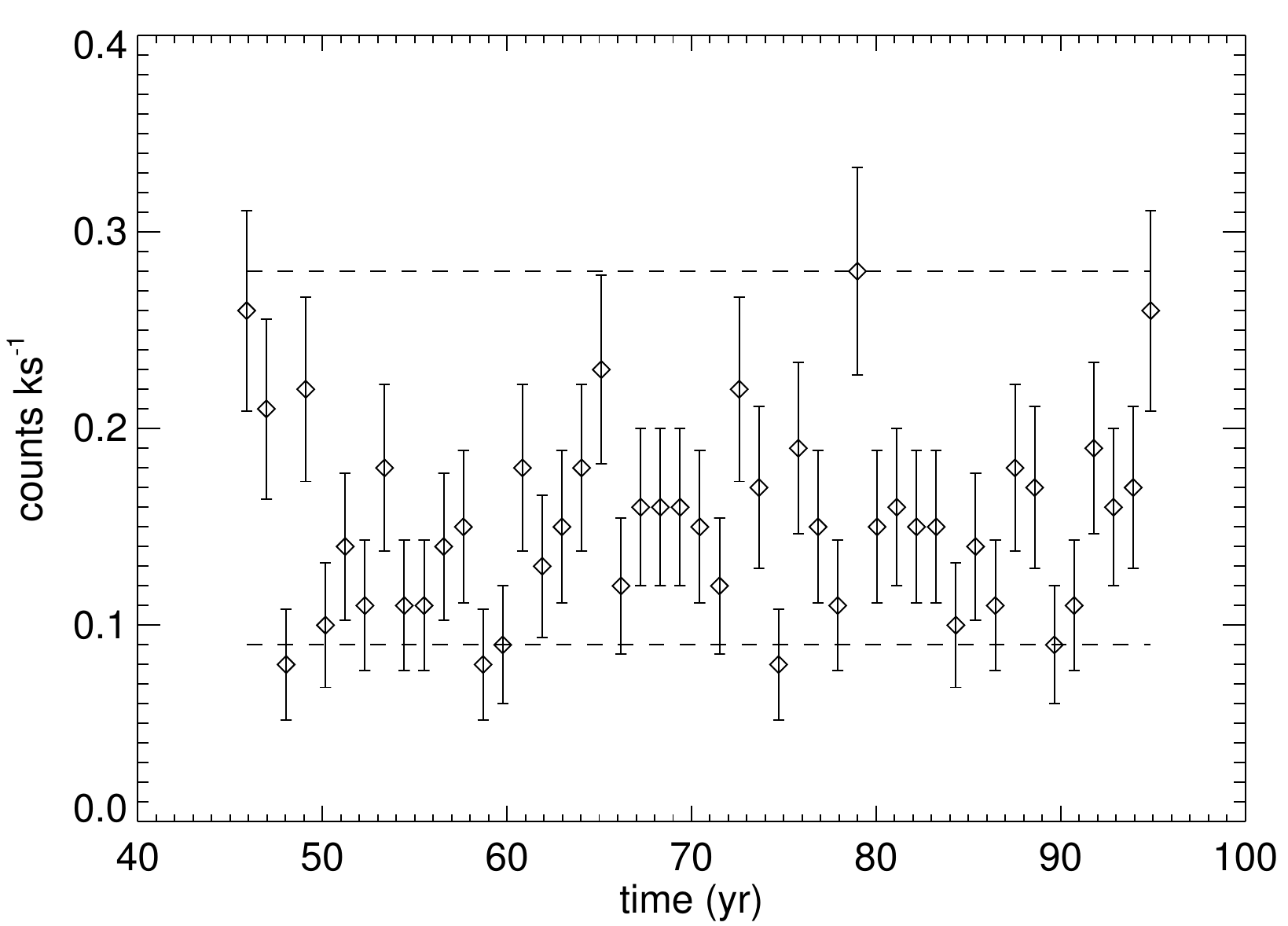}}
      \caption{X-ray count rate in the [0.5-1]~keV band with error bars for the 
      model HJ3. We plot one point every year. The dashed lines indicate the 
      interval of the count rate observed for the DG~Tau jet. Note the different 
      scale for the $y$-axis with respect to Fig.~\ref{sum_HJ}.}
      \label{CRt_HJ}
   \end{figure}

   \begin{figure}
      \resizebox{\hsize}{!}{\includegraphics*{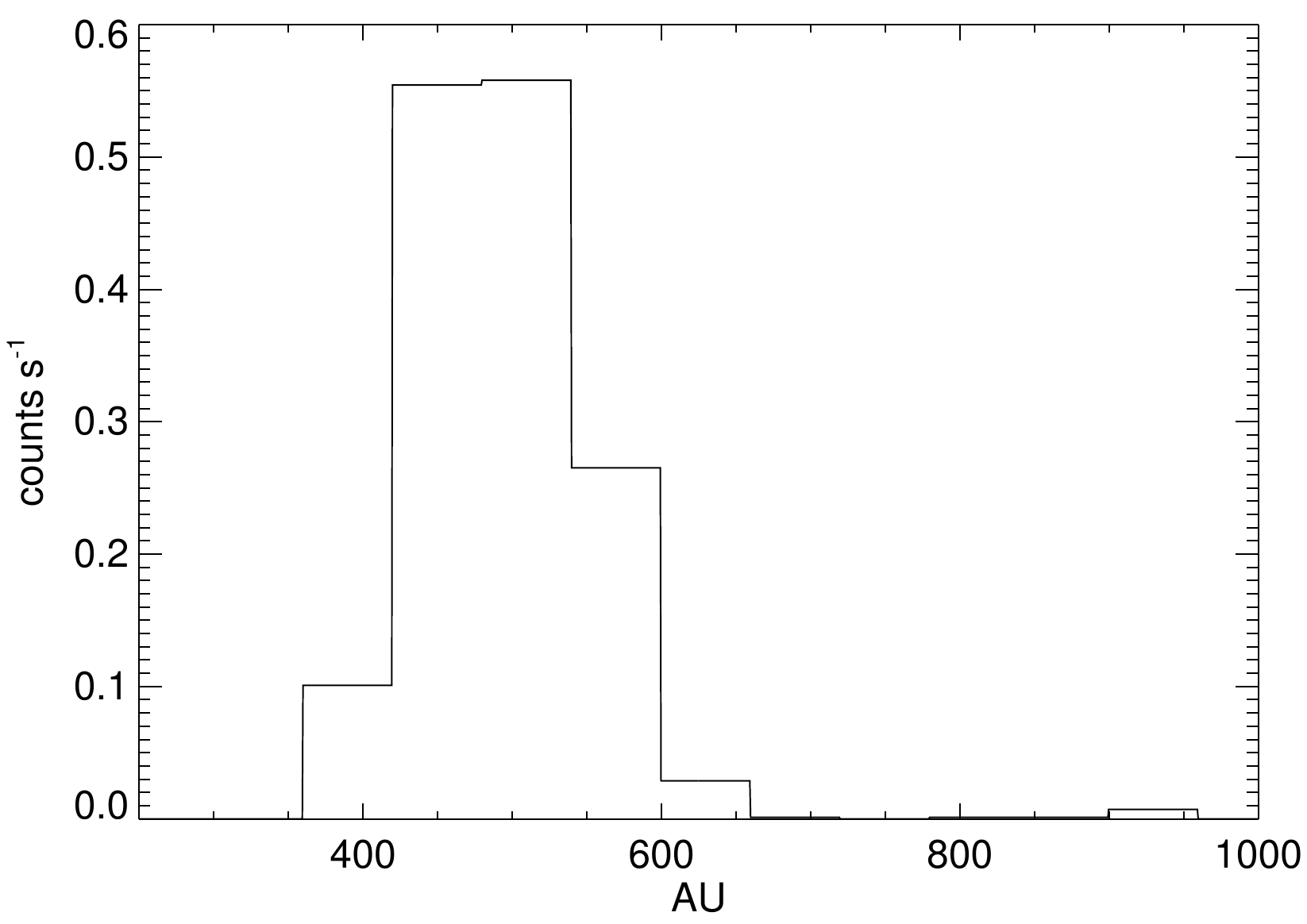}}
      \caption{X-ray count rate profile along the jet axis calculated integrating the 
      2D maps of the HJ3 model evolution in time and along the jet radius $r$.}
      \label{CRs2_HJ}
   \end{figure}

   \subsubsection{The reference case}
   
   As Fig.~\ref{res_LJ}, Fig.~\ref{res_HJ} reports the 2D spatial distributions 
   of temperature, density, and count rate for the model HJ3 
   (see Table~\ref{parameters}) in three different moments. The complete 
   temporal evolution is available as an online movie (Movie 2). The animation 
   starts when the pulsed component is introduced, formed by a train of blobs 
   with blob-to-jet particle number density ratio $\chi_{\mathrm{b}} = 1.5$ 
   (see Table~\ref{parameters}), and covers the evolution of the pulsed jet 
   for approximately 50 years.

   The left panels in Figure~\ref{res_HJ} show the stationary shock when 
   the first blob is arriving at $t \approx 42$~yr, for the model HJ3. The 
   stationary shock forms in the same way as described in Sect.~\ref{sec:LJ} 
   for the light jet scenario \citep[see also][]{ust16}.
   The plasma density and temperature reach respective maximum values 
   of $\sim 1\cdot10^{4}\,\mathrm{cm}^{-3}$ and 
   $\sim 5\cdot10^{6}\,\mathrm{K}$ at the shock. The shock temperature, 
   calculated as the density-weighted average temperature considering only 
   the cells with $T\geq 10^6$, is $\sim 2\cdot 10^6$~K. The pre-shock 
   density is $\sim 1000\,\mathrm{cm}^{-3}$, one order of magnitude higher
   than the ambient medium density, namely $\sim 100\,\mathrm{cm}^{-3}$.
   The count rate map in the [0.5-1]~keV band, calculated as described in 
   Section~\ref{sec:xray}, shows that the X-ray emission comes mainly from 
   the shock diamond (see lower left panel in Fig.~\ref{res_HJ}).
   The X-ray total shock luminosity, $L_{\mathrm{X}}$, derived in the 
   [0.5-1]~keV band is $\sim 2\cdot 10^{29}$~erg~s$^{-1}$.
   In Figure~\ref{plot_HJ} (upper panel) we plot the pressure profiles along 
   the jet axis at $t \approx 42$~yr. As in the light jet scenario, the model 
   evolution close to the jet axis is dominated by the jet plasma pressure 
   ($P$, represented in green) over the magnetic pressure 
   ($P_{\mathrm{m}}$, represented in blue), where $\beta>1$. 
   In this case the plasma pressure, and so on $\beta$, is lower comparing 
   with the case studied in the light jet scenario (see Fig.~\ref{plot_LJ}). 
   We plot in black the ram pressure, $P_{\mathrm{r}}$, and the dynamic 
   pressure, $H$, in red, which is almost constant 
   along the profile due to the stability and quasi-stationarity of the model. 
   The central panels in Figure~\ref{res_HJ} show the jet at $t \approx 44$~yr, 
   when the first blob just passed through the shock and the second one 
   is arriving. The shocked plasma density and temperature are higher after 
   the perturbation: maximum values of 
   $\sim 6\cdot10^{4}\,\mathrm{cm}^{-3}$ and $\sim 1\cdot10^{7}\,\mathrm{K}$,
   and density-weighted average temperature of $\sim 3\cdot 10^6$~K. 
   The X-ray source is perturbed but it is still located at the base of the jet 
   (see lower middle panel in Fig.~\ref{res_HJ}).
   The X-ray total shock luminosity, $L_{\mathrm{X}}$, in the [0.5-1]~keV band 
   is $\sim2\cdot 10^{29}$~erg~s$^{-1}$.
   In Figure~\ref{plot_HJ} (middle panel) we observe the perturbation effect in 
   the pressure profiles along the jet axis at $t \approx 44$~yr, when the blob 
   has just passed, affecting strongly the profiles stability.
   The right panels in Figure~\ref{res_HJ} show the jet at $t \approx 70$~yr, 
   after a train of blobs passed through the shock. The shocked plasma 
   maximum values for density and temperature are 
   $\sim 5\cdot10^{4}\,\mathrm{cm}^{-3}$ and $\sim 1\cdot10^{7}\,\mathrm{K}$ 
   respectively, and density-weighted average temperature of $\sim 2\cdot 10^6$~K. 
   The X-ray source is located at the base of the jet (see lower right panel in 
   Fig.~\ref{res_HJ}), with a total shock luminosity, $L_{\mathrm{X}}$, similar 
   to the values derived before, namely $\sim2\cdot 10^{29}$~erg~s$^{-1}$.
   In Figure~\ref{plot_HJ} (lower panel) we observe the pressure profiles along 
   the jet axis at $t \approx 70$~yr, completely perturbed by the train of blobs.

   \subsubsection{Variability}

   As in Sect.~3.1.2 we investigate the stationarity of the 
   different pulsed jet models by studying the variations of the physical 
   quantities and of the spatial distribution of the count rate during the 
   evolution. 
   We derive the total count rate in the [0.5-1]~keV band, calculated as 
   described in Sect.~\ref{sec:xray} and integrated in all the domain to obtain 
   the total value. 
   In Figure~\ref{CRt_HJ} we show one value of count rate with Poisson error 
   bars per year, omitting the first five values corresponding to the intial 
   transient of the pulsed jet observed at the begining of the animation (Movie 2).
   The dashed lines correpond to the interval [0.09,0.28], which contains 
   the count rate values derived from observations by 
   \citet{gud05,gud08,gud11}. We note that the values derived from the 
   model are in good agreement with the observations during the whole evolution 
   for the parameters explored here.\footnote{See Table~\ref{parameters} and 
   Sect.\ref{sec:description} for a complete description of the parameters explored.}
   The temporal evolution of the spatial distribution of the count rate for the 
   model HJ3, as would be seen by the \textit{Chandra}/ACIS instrument, 
   is reported in the right panel of the second movie (Movie 2). In this case, we 
   observe one single component emitting in X-rays at the base of the jet 
   throughout almost the whole animation. In order to understand the trend, 
   in the same way as described in Sect.~\ref{sec:LJ}, we derive the 
   count rate profile along the jet axis (see Fig.~\ref{CRs2_HJ}). We find that 
   most of the X-ray emission comes from one single source close to the 
   base of the jet.

   \subsubsection{Comparison with the other models}

   We compare different models for the heavy jet scenario in order 
   to investigate the effect of perturbations in the stationarity of the shock. 
   The explored parameters, describing the jet and the blobs, are listed 
   in Table~\ref{parameters}. The jet temperature and density, and the 
   possible rotation, determine the physical parameters of the shock 
   (see \citealt{ust16}). They are selected according to the observations 
   of DG Tau \citep{gud05,gud08}.
   The parameters of the blob (density, velocity and radius) and the mass 
   loss rate, calculated as explained in Section~\ref{sec:description}, give 
   information about the intensity of the perturbations introduced. 
   In our reference model (HJ3) the jet initial temperature and density are 
   $T_{\mathrm{j}}=10^{6}$~K and $n_{\mathrm{j}}=10^{4}$~cm$^{-3}$
   respectively. In cases with higher values of jet temperature and density, 
   namely HJ9-HJ12 models, the shock is stronger and the count rate is higher 
   (see Fig.~\ref{sum_HJ}). As in the light jet scenario, for the models 
   including jet rotation, namely HJ2, HJ3 and HJ6, the count rate is higher 
   because the shock is stronger (as predicted by \citealt{ust16}). This trend is 
   clear when we compare same models with and without jet rotation, 
   e.g. HJ6 and HJ4 (see Fig.~\ref{sum_HJ}). Again we do not 
   observe significant effect in the stability of the shock due to the rotation. 
   The parameters that define blob density and radius in the model HJ3 
   (reference case) are $\chi_{\mathrm{b}}=1.5$, and $R_{\mathrm{b}}=1/3$ 
   respectively, while the velocity $\varv_{\mathrm{b}}$ vary randomly from 
   300 to 700 km~s$^{-1}$ (see Table~\ref{parameters}). The random 
   velocity make the perturbation stronger due to the higher variations in 
   the ram pressure lowering slightly the count rate, although the effect in 
   the studied cases is very faint and not significant (see models HJ1 and 
   HJ4 in Fig.~\ref{sum_HJ}). 
   When the blob radius is higher (e.g. HJ2 with respect to HJ3, and HJ4 with 
   respect to HJ5) the shock is slightly affected lowering the median count rate 
   and enhancing the fluctuations. When the blob density is increased, we find
   different effects depending on the part of the domain where we are. In low
   count rate models an increase on the blob density perturbs the shock 
   lowering the count rate (e.g. HJ5 with respect to HJ8), while in high count 
   rate models the median count rate increases contributing to the emission
   (e.g. HJ12 with respect to HJ10), as in the ligh jet scenario. 
   We investigate the difference between the models through the pressure 
   profiles (see Figures~\ref{plot_LJ}~and~\ref{plot_HJ}). In all the cases the
   jet is dominated by the plasma pressure over the magnetic pressure, 
   i.e. $\beta>1$, but in cases with low count rate we find a more instable 
   regime because the $\beta$ parameter is closer to 1 in some moments of 
   the simulation. The latter makes that the perturbations affect in a different 
   way models with high and low count rate. The model HJ9 is situated on the 
   limit of the two different regimes described, and when higher density blobs
   are introduced (model HJ11), the median count rate remains almost constant.
   
   In summary, we can afirm that the  perturbations are compatible with the 
   available observations of DG~Tau in most of the cases explored. The jet 
   forms a quasi-stationary X-ray emitting shock at the base which is most 
   likely affected by perturbations arriving, as a train of blobs, from the stellar 
   source. In this case, the X-ray emitting shock results to be more perturbed 
   than that of HH~154 due to the low count rate observed for the DG Tau jet. 
   Strong perturbations can almost erase the X-ray emission from a shock 
   in models with low count rate, while in models with higher count rate 
   values the X-ray emission is enhanced, in a similar way to the light jet 
   scenario of HH~154.
   The variations registered in the count rate are comparable with those 
   observed for perturbations in the following ranges: density increase of 
   a maximum of three times the previous value, velocity fluctuation of ~50\% 
   and radius with values from 1/3-to-1 jet radius are compatible. The mass 
   loss rate values obtained for all the models are 
   $\dot M_{\mathrm{j}} \sim 10^{-8} M_{\odot}$yr$^{-1}$, in good agreement 
   with typical outflow rates found in pre-main sequence stars \citep{cab07,pod11}. 
   The maximum change in $\dot M_{\mathrm{j}}$ is approximately half order of 
   magnitude in the model HJ1, for which the total count rate is very low and it is 
   not compatible with the values observed for DG Tau by \cite{gud05,gud08,gud11}.

\clearpage

\section{Discussion}

   \begin{figure*}
      \resizebox{\hsize}{!}{\includegraphics*{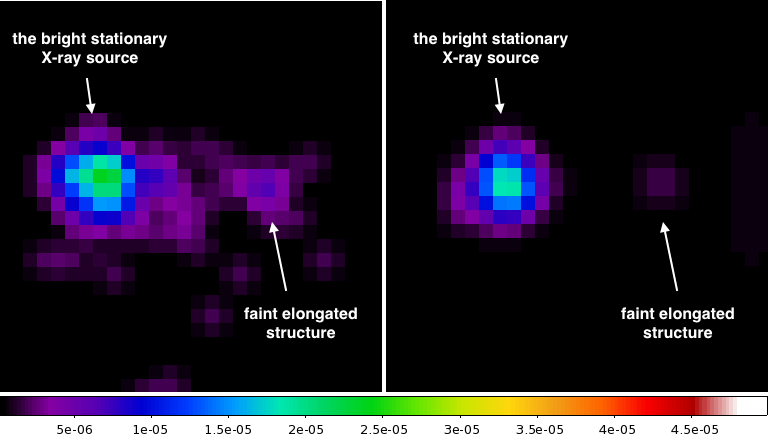}}
      \caption{Smoothed X-ray count rate maps in the [0.3–4]~keV band 
      for HH 154 with a pixel size of 0.25\arcsec. 
      On the left the 2005 data set resampled using EDSER technique as 
      in \citet{bon11}. 
      On the right synthetic image of the base of the jet as derived from the 
      model LJ5 at $t=100$~yr (see lower right panel in Fig.~\ref{res_LJ}), 
      rebinned to match the same pixel size and convolved with the proper PSF.  
      The angular size of each panel is $\approx 7\arcsec \times 7\arcsec$. 
      In each panel north is up and east is left.
      Gaussian smoothing was applied on the images with kernel of width 
      0.5\arcsec.}
      \label{HH154}
   \end{figure*}
   
   \begin{figure*}
      \resizebox{\hsize}{!}{\includegraphics*{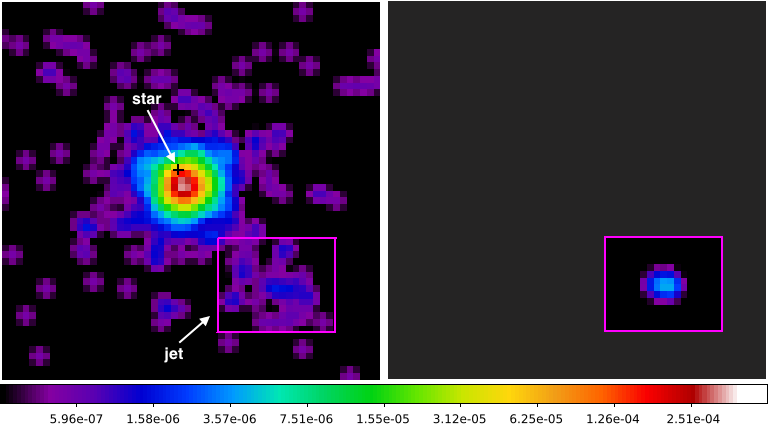}}
      \caption{Smoothed X-ray count rate maps in the [0.5–1]~keV band 
      for DG Tau in logarithmic scale and with a pixel size of 0.25\arcsec. 
      On the left the 2010 data set resampled using EDSER technique as 
      explained in Appendix A. We marked the position of the star with a 
      cross and the SW jet with a box.
      On the right synthetic image of the base of the jet as derived from the 
      model HJ3 at $t=70$~yr (see lower right panel in Fig.~\ref{res_HJ}), 
      rebinned to match the same pixel size and convolved with the specific PSF.  
      The modeled region is marked with a box corresponding to the SW jet.
      The angular size of each panel is $\approx 14\arcsec \times 14\arcsec$
      (note the different scale with respect to Fig.~\ref{HH154}). 
      In each panel north is up and east is left.
      Gaussian smoothing was applied on the images with kernel of width 
      0.5\arcsec.}
      \label{DGTAU}
   \end{figure*}
   
   In a previous study \citep{ust16}, we showed that a continuously driven 
   stellar jet forms a stationary X-ray emitting shock at the base with physical 
   parameters in good agreement with observations. The aim of this work is 
   to investigate if the quasi-stationary shocks formed are compatible with 
   the perturbations expected in YSO jets through simulations of a pulsed flow, 
   and derive the physical parameters that can give rise to X-ray emission 
   consistent with observations of jets in pre-main sequence stars. To this end
   we developed a magnetohydrodynamic 2.5-dimensional model that describes 
   the propagation of a supersonic stellar jet in a initially homogeneous 
   magnetized medium taking into account the effect of the radiative cooling. 
   The jet is described by two components: a continously driven component 
   that forms a quasi-stationary shock at the base of the jet; and a pulsed 
   component formed by blobs that introduce perturbations into the jet 
   influencing the shock.
   In order to compare the model results with observations we 
   synthesized the count rate from the simulations, considering both its 
   total integrated value and its spatial distribution.
   
   Previous MHD models of protostellar jets were more oriented at studying 
   the dynamical aspects and the evolution of the jet rather than on the X-ray 
   emission produced. Here we have shown the feasibility of our 
   MHD model: a supersonic stellar jet collimated by the ambient 
   magnetic field leading to X-ray emission from the shock formed at the base, 
   consistent with the observations. 
   We obtained shock temperatures of $\sim 10^6$~K and 
   luminosities of $L_\mathrm{X} \approx 10^{29}$~erg~s$^{-1}$, in good 
   agreement with the X-ray results of other authors 
   \citep{pra01,fav02,bal03,pra05,gud08, sch11,ski11,lop13,lop15}.
      
   As YSO jets are intrinsically dynamic objects which evolve on timescales 
   of a few years, they are ideal laboratories for variability studies. We selected 
   HH~154 and DG~Tau as specific targets for our study because available 
   observations suggest the presence of stationary X-ray emitting sources 
   close to the base associated with the jet 
   (HH~154, \citealt{fav06}; DG~Tau, \citealt{gud05}).
   \textit{Chandra} X-ray observations collected in different epochs for HH~154  
   (in 2001, \citealt{bal03}; in 2005, \citealt{fav06}; in 2009, \citealt{sch11}) and 
   for DG~Tau's jet \citep[in 2004, 2006 and 2010,][]{gud05,gud08,gud11}, 
   provided a time base of eight and six years respectively to analyze the 
   variability of the source. 
   They are, as far as we know, the only YSO jets with available multi-epoch 
   X-ray observations.

\subsection{The case of HH~154}
   
   In the light jet scenario (see Section~\ref{sec:LJ}) we described a jet less dense 
   than the ambient medium, as the jet related to HH~154 which originates from 
   the deeply embedded binary Class 0/I protostar IRS 5 in the L1551 star-forming 
   region \citep{rod98}.
   
   Figure~\ref{HH154} shows the count rate of the X-ray source associated 
   with HH 154 in the [0.3–4]~keV band. We compare the data set observed 
   by \textit{Chandra}/ACIS in 2005 and analysed by \citet{bon11} (left panel), 
   with the synthetic map derived from the model LJ5 (reference case for the 
   light jet presented in Section~\ref{sec:LJ}) at $t=100$~yr, and convolved 
   with the specific PSF created at the proper energy (right panel). The 
   images are rebinned to match a pixel size of 0.25\arcsec and then a 
   Gaussian smoothing with kernel of width 0.5\arcsec was applied. 
   \citet{bon11} analysed Chandra multi-epoch X-ray observations of HH~154 
   (in 2001, \citealt{bal03}; in 2005, \citealt{fav06}; in 2009, \citealt{sch11}) 
   and proposed the scenario of a nozzle creating the standing shock, 
   in the presence of a pulsed jet, as described in \citet{bon10a}, which may 
   account for the elongated component as a newly formed knot propagating 
   away from the diamond shock. Here we propose an alternative possible 
   interpretation: the bright stationary X-ray source observed in the data 
   might be associated to the main stationary shock formed at the base 
   of the jet, whereas the faint, elongated and more variable component 
   could be the result of perturbations of the main shock. 
   As discussed by \citet{bon11}, the X-ray source associated to HH~154 
   unambiguously arises from the jet and cannot be of stellar origin.

\subsection{The case of the jet associated to DG~Tau}
   
   In the heavy jet scenario (see Section~\ref{sec:HJ}) we described a jet denser
   than the ambient medium, as the jet associated with DG Tau which originates 
   from a more evolved Class II disk-bearing source (CTTS) \cite[see][]{eis98}. 
   
   In Figure~\ref{DGTAU} we show the count rate of the X-ray source associated 
   with DG Tau in the [0.5–1]~keV band, in logarithmic scale. On the left panel we 
   plot the merged data set observed by \textit{Chandra}/ACIS in 2010 
   (see Appendix A for a detailed description of the data analysis we performed) 
   with a pixel size of 0.25\arcsec and Gaussian smoothing with kernel of width 
   0.5\arcsec. In this case the angular size of each panel is 
   $\approx 14\arcsec \times 14\arcsec$ (note the different scale with respect to 
   Fig.~\ref{HH154} which corresponds only to one quadrant of this image). 
   The observation shows the star marked with a cross inside the central 
   unresolved source, and a bipolar jet. In this image the SW jet marked 
   with a box is clearly visible while the counter-jet (observed before by 
   \citealt{gud05,gud08}) is hardly visible. The central unresolved source contains 
   two unrelated spectral components subject to different hydrogen absorption 
   column densities \citep{gud11}: the hard component is ascribed to the stellar 
   corona or magnetosphere, while the soft component is associated with X-ray 
   emission from the base of the jet produced either by internal shocks or by 
   magnetic heating \citep{gud08}. Recently, \citet{tak17} presented a theoretical 
   model applied to DG~Tau in which the disk atmosphere is magnetically heated 
   forming a hot corona emitting in soft X-rays. However, in this data it is still unclear 
   which part of the soft component of the X-ray emission is produced by the jet due to 
   the contamination from the star. For this reason, we compare our jet results with 
   the SW jet marked with a box in the left panel in Figure~\ref{DGTAU}. On the
   right we plot the synthetic map of the modeled region derived from the 
   model HJ3 (reference case for the heavy jet presented in Section~\ref{sec:HJ}) 
   at $t=70$~yr, and convolved with the specific PSF created at the proper energy. 
   Again the image is rebinned to match a pixel size of 0.25\arcsec and then 
   smoothed with a Gaussian kernel of width 0.5\arcsec. In this case the 
   synthesized source consists of one unique component, while in the case of 
   HH~154 (see Fig.~\ref{HH154}) we could observe cleary two different components.
   The size and morphology of the synthesized source is comparable with 
   the SW jet observations.

      
\section{Summary and conclusions}

   We investigated the effect of perturbations in X-ray emitting stationary 
   shocks in stellar jets, throug numerical MHD simulations.
   We applied our model to the X-ray jets of HH~154 and DG~Tau, two widely 
   studied objects at different stages of evolution. This allowed us to explore 
   the similarities and differences of YSOs at distinct stages of evolution, 
   through the study of the X-ray emission and jets which are present in objects 
   from Class~0~to~II. 
   We have also performed a wide exploration of a broad region of the parameter 
   space that describes the model (see Table~\ref{parameters}). These results
   therefore allow us to study and diagnose the physical properties of YSO jets over 
   a broader range of physical conditions than those defined by HH~154 and DG~Tau.
   Our findings lead to several conclusions, that we list in the following for the two 
   different scenario studied (LJ and HJ).\\
   
   \begin{itemize}
      \item LJ scenario (HH~154):
      \begin{enumerate}   
      \item We find that perturbations arriving to the shock as a 
      train of blobs contribute to the X-ray emission enhancing the total 
      count rate.
      \item Perturbations characterized by a density increase up 
      to one order of magnitude, velocity fluctuations not larger than 50\%, 
      and size with values from 1/3-to-1 of the initial jet radius produce 
      variability of X-ray source compatible with those observed in HH~154 
      \citep{bon11}.
      \item The perturbations explored lead to maximum variations 
      of approximately one order of magnitude in the mass loss rate 
      derived from the simulations.\\
      \end{enumerate}
      
      \item HJ scenario (DG~Tau):
      \begin{enumerate} 
      \item The stability of the shock diamond is affected by 
      the jet perturbations more easily in the HJ than in the LJ scenario. 
      This is mainly due to the fact that the shock in the HJ scenario is 
      fainter and with a lower total count rate than in the LJ scenario 
      (see Fig.~\ref{sum_HJ}). In addition, models with a lower median 
      count rate (e.g. HJ1-HJ8) are more affected by the perturbations 
      than models with higher median count rate (e.g. HJ9-HJ12). 
      In the latter models, the perturbations enhance the X-ray emission 
      as in the light jet scenario.
      \item Perturbations characterized by a density increase up 
      to three times the initial jet value, velocity fluctuations not larger 
      than 50\%, and size with values from 1/3-to-1 initial jet radius produce 
      variability of X-ray source compatible with those observed in the 
      SW jet of DG~Tau \citep{gud08,gud11}.
      \item The perturbations explored lead to maximum variations 
      of $\sim 1-4 \cdot \dot M_{\mathrm{j}}$ in the mass loss rate derived 
      from the simulations (see Table~\ref{parameters}), corresponding to 
      fainter perturbations than those presented in the LJ scenario.\\
      \end{enumerate} 
   \end{itemize}

   In both the scenarios explored, although the pysical conditions are very 
   different, the plasma is collimated by the magnetic field forming a 
   quasi-stationary shock at the base of the jet which, under certain conditions, 
   emits in X-rays even when perturbations are present. 
   The results presented here may allow us a better understanding of the evolution 
   and the different mechanisms observed in young stars at different stages of 
   evolution. Although the precise mechanisms are still under debate, it is widely 
   believed that the strongly dynamic and energetic phenomena due to star-disk 
   interaction produce variations in the physical parameters of the jet.
   The study of the variability observed in pre-main sequence stars may give 
   some insight into the phenomena that occur due to the star disk interaction, 
   also related to the still debated jet collimation and acceleration mechanisms. 
   Finally, the comparison of our MHD model results with the X-ray observations 
   could provide a fundamental tool for investigating the stellar jet dynamics and 
   the high-energy phenomena, also important to better understanding the planet 
   formation.


\begin{acknowledgements}

   S.U. acknowledges the hospitality of the INAF Osservatorio Astronomico 
   di Palermo, where part of the present work was carried out using the HPC 
   facility (SCAN).
   PLUTO is being developed at the Turin Astronomical Observatory in 
   collaboration with the Department of General Physics of the Turin University.
   This research has made use of data obtained from the \textit{Chandra} 
   Data Archive (published previously in cited articles), and software provided 
   by the \textit{Chandra} X-ray Center (CXC) in the application package CIAO.
   This work was supported by grant BES-2012-061750 from the Spanish 
   Government under research project AYA2011-29754-C03-01, and by the 
   Ministry of Economy and Competitivity of Spain under grant numbers 
   ESP2014-54243-R and ESP2015-68908-R.
   The research leading to these results has also received funding from the 
   European Union’s Horizon 2020 Programme under the AHEAD project 
   (grant agreement n. 654215).
   R. B. acknowledges financial support from INAF under PRIN2013 
   Programme 'Disks, jets and the dawn of planets'.
   We would like to thank Dr. Francesco Damiani for helpful discussions.
   Finally, we thank the referee for useful comments and suggestions.
   
\end{acknowledgements}


\bibliographystyle{aa} 
\bibliography{biblio.bib} 

\begin{thebibliography}{67}
\expandafter\ifx\csname natexlab\endcsname\relax\def\natexlab#1{#1}\fi

\bibitem[{{Albertazzi} {et~al.}(2014){Albertazzi}, {Ciardi}, {Nakatsutsumi},
  {Vinci}, {B{\'e}ard}, {Bonito}, {Billette}, {Borghesi}, {Burkley}, {Chen},
  {Cowan}, {Herrmannsd{\"o}rfer}, {Higginson}, {Kroll}, {Pikuz}, {Naughton},
  {Romagnani}, {Riconda}, {Revet}, {Riquier}, {Schlenvoigt}, {Skobelev},
  {Faenov}, {Soloviev}, {Huarte-Espinosa}, {Frank}, {Portugall}, {P{\'e}pin},
  \& {Fuchs}}]{alb14}
{Albertazzi}, B., {Ciardi}, A., {Nakatsutsumi}, M., {et~al.} 2014, Science,
  346, 325

\bibitem[{{Anders} \& {Grevesse}(1989)}]{and89}
{Anders}, E. \& {Grevesse}, N. 1989, \gca, 53, 197

\bibitem[{{Andre} \& {Montmerle}(1994)}]{and94}
{Andre}, P. \& {Montmerle}, T. 1994, \apj, 420, 837

\bibitem[{{Bacciotti} {et~al.}(2002){Bacciotti}, {Ray}, {Mundt},
  {Eisl{\"o}ffel}, \& {Solf}}]{bac02}
{Bacciotti}, F., {Ray}, T.~P., {Mundt}, R., {Eisl{\"o}ffel}, J., \& {Solf}, J.
  2002, \apj, 576, 222

\bibitem[{{Bally} {et~al.}(2003){Bally}, {Feigelson}, \& {Reipurth}}]{bal03}
{Bally}, J., {Feigelson}, E., \& {Reipurth}, B. 2003, \apj, 584, 843

\bibitem[{{Balsara} \& {Spicer}(1999)}]{bal99}
{Balsara}, D.~S. \& {Spicer}, D.~S. 1999, Journal of Computational Physics,
  149, 270

\bibitem[{{Bodo} {et~al.}(1994){Bodo}, {Massaglia}, {Ferrari}, \&
  {Trussoni}}]{bod94}
{Bodo}, G., {Massaglia}, S., {Ferrari}, A., \& {Trussoni}, E. 1994, \aap, 283,
  655

\bibitem[{{Bonito} {et~al.}(2008){Bonito}, {Fridlund}, {Favata}, {Micela},
  {Peres}, {Djupvik}, \& {Liseau}}]{bon08}
{Bonito}, R., {Fridlund}, C.~V.~M., {Favata}, F., {et~al.} 2008, \aap, 484, 389

\bibitem[{{Bonito} {et~al.}(2010{\natexlab{a}}){Bonito}, {Orlando}, {Miceli},
  {Eisl{\"o}ffel}, {Peres}, \& {Favata}}]{bon10b}
{Bonito}, R., {Orlando}, S., {Miceli}, M., {et~al.} 2010{\natexlab{a}}, \aap,
  517, A68

\bibitem[{{Bonito} {et~al.}(2011){Bonito}, {Orlando}, {Miceli}, {Peres},
  {Micela}, \& {Favata}}]{bon11}
{Bonito}, R., {Orlando}, S., {Miceli}, M., {et~al.} 2011, \apj, 737, 54

\bibitem[{{Bonito} {et~al.}(2010{\natexlab{b}}){Bonito}, {Orlando}, {Peres},
  {Eisl{\"o}ffel}, {Miceli}, \& {Favata}}]{bon10a}
{Bonito}, R., {Orlando}, S., {Peres}, G., {et~al.} 2010{\natexlab{b}}, \aap,
  511, A42

\bibitem[{{Bonito} {et~al.}(2004){Bonito}, {Orlando}, {Peres}, {Favata}, \&
  {Rosner}}]{bon04}
{Bonito}, R., {Orlando}, S., {Peres}, G., {Favata}, F., \& {Rosner}, R. 2004,
  \aap, 424, L1

\bibitem[{{Bonito} {et~al.}(2007){Bonito}, {Orlando}, {Peres}, {Favata}, \&
  {Rosner}}]{bon07}
{Bonito}, R., {Orlando}, S., {Peres}, G., {Favata}, F., \& {Rosner}, R. 2007,
  \aap, 462, 645

\bibitem[{{Cabrit}(2007)}]{cab07_LNP}
{Cabrit}, S. 2007, in Lecture Notes in Physics, Berlin Springer Verlag, Vol.
  723, Lecture Notes in Physics, Berlin Springer Verlag, ed. J.~{Ferreira},
  C.~{Dougados}, \& E.~{Whelan}, 21

\bibitem[{{Cabrit} {et~al.}(2007){Cabrit}, {Codella}, {Gueth}, {Nisini},
  {Gusdorf}, {Dougados}, \& {Bacciotti}}]{cab07}
{Cabrit}, S., {Codella}, C., {Gueth}, F., {et~al.} 2007, \aap, 468, L29

\bibitem[{{Cabrit} {et~al.}(1990){Cabrit}, {Edwards}, {Strom}, \&
  {Strom}}]{cab90}
{Cabrit}, S., {Edwards}, S., {Strom}, S.~E., \& {Strom}, K.~M. 1990, \apj, 354,
  687

\bibitem[{{Ciardi} {et~al.}(2009){Ciardi}, {Lebedev}, {Frank}, {Suzuki-Vidal},
  {Hall}, {Bland}, {Harvey-Thompson}, {Blackman}, \& {Camenzind}}]{cia09}
{Ciardi}, A., {Lebedev}, S.~V., {Frank}, A., {et~al.} 2009, \apjl, 691, L147

\bibitem[{{Coffey} {et~al.}(2007){Coffey}, {Bacciotti}, {Ray}, {Eisl{\"o}ffel},
  \& {Woitas}}]{cof07}
{Coffey}, D., {Bacciotti}, F., {Ray}, T.~P., {Eisl{\"o}ffel}, J., \& {Woitas},
  J. 2007, \apj, 663, 350

\bibitem[{{de Colle} \& {Raga}(2006)}]{col06}
{de Colle}, F. \& {Raga}, A.~C. 2006, \aap, 449, 1061

\bibitem[{{Eisl{\"o}ffel} \& {Mundt}(1998)}]{eis98}
{Eisl{\"o}ffel}, J. \& {Mundt}, R. 1998, \aj, 115, 1554

\bibitem[{{Favata} {et~al.}(2006){Favata}, {Bonito}, {Micela}, {Fridlund},
  {Orlando}, {Sciortino}, \& {Peres}}]{fav06}
{Favata}, F., {Bonito}, R., {Micela}, G., {et~al.} 2006, \aap, 450, L17

\bibitem[{{Favata} {et~al.}(2002){Favata}, {Fridlund}, {Micela}, {Sciortino},
  \& {Kaas}}]{fav02}
{Favata}, F., {Fridlund}, C.~V.~M., {Micela}, G., {Sciortino}, S., \& {Kaas},
  A.~A. 2002, \aap, 386, 204

\bibitem[{{Feigelson} \& {Montmerle}(1999)}]{fei99}
{Feigelson}, E.~D. \& {Montmerle}, T. 1999, \araa, 37, 363

\bibitem[{{Ferreira} {et~al.}(2006){Ferreira}, {Dougados}, \& {Cabrit}}]{fer06}
{Ferreira}, J., {Dougados}, C., \& {Cabrit}, S. 2006, \aap, 453, 785

\bibitem[{{Frank} {et~al.}(2014){Frank}, {Ray}, {Cabrit}, {Hartigan}, {Arce},
  {Bacciotti}, {Bally}, {Benisty}, {Eisl{\"o}ffel}, {G{\"u}del}, {Lebedev},
  {Nisini}, \& {Raga}}]{fra14}
{Frank}, A., {Ray}, T.~P., {Cabrit}, S., {et~al.} 2014, Protostars and Planets
  VI, 451

\bibitem[{{Fridlund} {et~al.}(1998){Fridlund}, {Liseau}, \&
  {Gullbring}}]{fri98}
{Fridlund}, C.~V.~M., {Liseau}, R., \& {Gullbring}, E. 1998, \aap, 330, 327

\bibitem[{{G{\"u}del} {et~al.}(2011){G{\"u}del}, {Audard}, {Bacciotti}, {Bary},
  {Briggs}, {Cabrit}, {Carmona}, {Codella}, {Dougados}, {Eisl{\"o}ffel},
  {Gueth}, {G{\"u}nther}, {Herczeg}, {Kundurthy}, {Matt}, {Mutel}, {Ray},
  {Schmitt}, {Schneider}, {Skinner}, \& {van Boekel}}]{gud11}
{G{\"u}del}, M., {Audard}, M., {Bacciotti}, F., {et~al.} 2011, in Astronomical
  Society of the Pacific Conference Series, Vol. 448, 16th Cambridge Workshop
  on Cool Stars, Stellar Systems, and the Sun, ed. C.~{Johns-Krull}, M.~K.
  {Browning}, \& A.~A. {West}, 617

\bibitem[{{G{\"u}del} {et~al.}(2008){G{\"u}del}, {Skinner}, {Audard}, {Briggs},
  \& {Cabrit}}]{gud08}
{G{\"u}del}, M., {Skinner}, S.~L., {Audard}, M., {Briggs}, K.~R., \& {Cabrit},
  S. 2008, \aap, 478, 797

\bibitem[{{G{\"u}del} {et~al.}(2005){G{\"u}del}, {Skinner}, {Briggs}, {Audard},
  {Arzner}, \& {Telleschi}}]{gud05}
{G{\"u}del}, M., {Skinner}, S.~L., {Briggs}, K.~R., {et~al.} 2005, \apjl, 626,
  L53

\bibitem[{{Hartigan} {et~al.}(1995){Hartigan}, {Edwards}, \&
  {Ghandour}}]{har95}
{Hartigan}, P., {Edwards}, S., \& {Ghandour}, L. 1995, \apj, 452, 736

\bibitem[{{Kashyap} \& {Drake}(2000)}]{kas00}
{Kashyap}, V. \& {Drake}, J.~J. 2000, Bulletin of the Astronomical Society of
  India, 28, 475

\bibitem[{{Lada}(1987)}]{lad87}
{Lada}, C.~J. 1987, in IAU Symposium, Vol. 115, Star Forming Regions, ed.
  M.~{Peimbert} \& J.~{Jugaku}, 1--17

\bibitem[{{Li} {et~al.}(2004){Li}, {Kastner}, {Prigozhin}, {Schulz},
  {Feigelson}, \& {Getman}}]{lij04}
{Li}, J., {Kastner}, J.~H., {Prigozhin}, G.~Y., {et~al.} 2004, \apj, 610, 1204

\bibitem[{{L{\'o}pez-Santiago} {et~al.}(2015){L{\'o}pez-Santiago}, {Bonito},
  {Orellana}, {Miceli}, {Orlando}, {Ustamujic}, {Albacete-Colombo}, {de
  Castro}, \& {G{\'o}mez de Castro}}]{lop15}
{L{\'o}pez-Santiago}, J., {Bonito}, R., {Orellana}, M., {et~al.} 2015, \apj,
  806, 53

\bibitem[{{L{\'o}pez-Santiago} {et~al.}(2013){L{\'o}pez-Santiago}, {Peri},
  {Bonito}, {Miceli}, {Albacete-Colombo}, {Benaglia}, \& {de Castro}}]{lop13}
{L{\'o}pez-Santiago}, J., {Peri}, C.~S., {Bonito}, R., {et~al.} 2013, \apjl,
  776, L22

\bibitem[{{Masciadri} {et~al.}(2002){Masciadri}, {Vel{\'a}zquez}, {Raga},
  {Cant{\'o}}, \& {Noriega-Crespo}}]{mas02}
{Masciadri}, E., {Vel{\'a}zquez}, P.~F., {Raga}, A.~C., {Cant{\'o}}, J., \&
  {Noriega-Crespo}, A. 2002, \apj, 573, 260

\bibitem[{{Mignone} {et~al.}(2007){Mignone}, {Bodo}, {Massaglia}, {Matsakos},
  {Tesileanu}, {Zanni}, \& {Ferrari}}]{mig07}
{Mignone}, A., {Bodo}, G., {Massaglia}, S., {et~al.} 2007, \apjs, 170, 228

\bibitem[{{Orlando} {et~al.}(2011){Orlando}, {Reale}, {Peres}, \&
  {Mignone}}]{orl11}
{Orlando}, S., {Reale}, F., {Peres}, G., \& {Mignone}, A. 2011, \mnras, 415,
  3380

\bibitem[{{Podio} {et~al.}(2011){Podio}, {Eisl{\"o}ffel}, {Melnikov}, {Hodapp},
  \& {Bacciotti}}]{pod11}
{Podio}, L., {Eisl{\"o}ffel}, J., {Melnikov}, S., {Hodapp}, K.~W., \&
  {Bacciotti}, F. 2011, \aap, 527, A13

\bibitem[{{Pravdo} {et~al.}(2001){Pravdo}, {Feigelson}, {Garmire}, {Maeda},
  {Tsuboi}, \& {Bally}}]{pra01}
{Pravdo}, S.~H., {Feigelson}, E.~D., {Garmire}, G., {et~al.} 2001, \nat, 413,
  708

\bibitem[{{Pravdo} \& {Tsuboi}(2005)}]{pra05}
{Pravdo}, S.~H. \& {Tsuboi}, Y. 2005, \apj, 626, 272

\bibitem[{{Pravdo} {et~al.}(2004){Pravdo}, {Tsuboi}, \& {Maeda}}]{pra04}
{Pravdo}, S.~H., {Tsuboi}, Y., \& {Maeda}, Y. 2004, \apj, 605, 259

\bibitem[{{Pudritz} \& {Norman}(1983)}]{pud83}
{Pudritz}, R.~E. \& {Norman}, C.~A. 1983, \apj, 274, 677

\bibitem[{{Pudritz} \& {Norman}(1986)}]{pud86}
{Pudritz}, R.~E. \& {Norman}, C.~A. 1986, \apj, 301, 571

\bibitem[{{Pudritz} {et~al.}(2007){Pudritz}, {Ouyed}, {Fendt}, \&
  {Brandenburg}}]{pud07}
{Pudritz}, R.~E., {Ouyed}, R., {Fendt}, C., \& {Brandenburg}, A. 2007,
  Protostars and Planets V, 277

\bibitem[{{Raga} {et~al.}(2001){Raga}, {Cabrit}, {Dougados}, \&
  {Lavalley}}]{rag01}
{Raga}, A., {Cabrit}, S., {Dougados}, C., \& {Lavalley}, C. 2001, \aap, 367,
  959

\bibitem[{{Raga} \& {Noriega-Crespo}(1998)}]{rag98}
{Raga}, A. \& {Noriega-Crespo}, A. 1998, \aj, 116, 2943

\bibitem[{{Raga} {et~al.}(1990){Raga}, {Binette}, {Canto}, \& {Calvet}}]{rag90}
{Raga}, A.~C., {Binette}, L., {Canto}, J., \& {Calvet}, N. 1990, \apj, 364, 601

\bibitem[{{Raga} {et~al.}(2007){Raga}, {de Colle}, {Kajdi{\v c}}, {Esquivel},
  \& {Cant{\'o}}}]{rag07}
{Raga}, A.~C., {de Colle}, F., {Kajdi{\v c}}, P., {Esquivel}, A., \&
  {Cant{\'o}}, J. 2007, \aap, 465, 879

\bibitem[{{Raga} {et~al.}(2010){Raga}, {Riera}, \&
  {Gonz{\'a}lez-G{\'o}mez}}]{rag10}
{Raga}, A.~C., {Riera}, A., \& {Gonz{\'a}lez-G{\'o}mez}, D.~I. 2010, \aap, 517,
  A20

\bibitem[{{Raga} {et~al.}(2004){Raga}, {Riera}, {Masciadri}, {Beck},
  {B{\"o}hm}, \& {Binette}}]{rag04}
{Raga}, A.~C., {Riera}, A., {Masciadri}, E., {et~al.} 2004, \aj, 127, 1081

\bibitem[{{Reipurth} \& {Bally}(2001)}]{rei01}
{Reipurth}, B. \& {Bally}, J. 2001, \araa, 39, 403

\bibitem[{{Rodr{\'{\i}}guez} {et~al.}(1998){Rodr{\'{\i}}guez}, {D'Alessio},
  {Wilner}, {Ho}, {Torrelles}, {Curiel}, {G{\'o}mez}, {Lizano}, {Pedlar},
  {Cant{\'o}}, \& {Raga}}]{rod98}
{Rodr{\'{\i}}guez}, L.~F., {D'Alessio}, P., {Wilner}, D.~J., {et~al.} 1998,
  \nat, 395, 355

\bibitem[{{Romanova} {et~al.}(2011){Romanova}, {Long}, {Lamb}, {Kulkarni}, \&
  {Donati}}]{rom11}
{Romanova}, M.~M., {Long}, M., {Lamb}, F.~K., {Kulkarni}, A.~K., \& {Donati},
  J.-F. 2011, \mnras, 411, 915

\bibitem[{{Schneider} {et~al.}(2011){Schneider}, {G{\"u}nther}, \&
  {Schmitt}}]{sch11}
{Schneider}, P.~C., {G{\"u}nther}, H.~M., \& {Schmitt}, J.~H.~M.~M. 2011, \aap,
  530, A123

\bibitem[{{Skinner} {et~al.}(2011){Skinner}, {Audard}, \& {G{\"u}del}}]{ski11}
{Skinner}, S.~L., {Audard}, M., \& {G{\"u}del}, M. 2011, \apj, 737, 19

\bibitem[{{Smith} {et~al.}(2001){Smith}, {Brickhouse}, {Liedahl}, \&
  {Raymond}}]{smi01}
{Smith}, R.~K., {Brickhouse}, N.~S., {Liedahl}, D.~A., \& {Raymond}, J.~C.
  2001, \apjl, 556, L91

\bibitem[{{Stelzer}(2015)}]{ste15}
{Stelzer}, B. 2015, Astronomische Nachrichten, 336, 493

\bibitem[{{Stelzer}(2017)}]{ste17}
{Stelzer}, B. 2017, Astronomische Nachrichten, 338, 195

\bibitem[{{Stelzer} {et~al.}(2009){Stelzer}, {Hubrig}, {Orlando}, {Micela},
  {Mikul{\'a}{\v s}ek}, \& {Sch{\"o}ller}}]{ste09}
{Stelzer}, B., {Hubrig}, S., {Orlando}, S., {et~al.} 2009, \aap, 499, 529

\bibitem[{{Stone} {et~al.}(1996){Stone}, {Hawley}, {Gammie}, \&
  {Balbus}}]{sto96}
{Stone}, J.~M., {Hawley}, J.~F., {Gammie}, C.~F., \& {Balbus}, S.~A. 1996,
  \apj, 463, 656

\bibitem[{{Stone} \& {Norman}(1994)}]{sto94}
{Stone}, J.~M. \& {Norman}, M.~L. 1994, \apj, 433, 746

\bibitem[{{Takasao} {et~al.}(2017){Takasao}, {Suzuki}, \& {Shibata}}]{tak17}
{Takasao}, S., {Suzuki}, T.~K., \& {Shibata}, K. 2017, \apj, 847, 46

\bibitem[{{Telleschi} {et~al.}(2007){Telleschi}, {G{\"u}del}, {Briggs},
  {Skinner}, {Audard}, \& {Franciosini}}]{tel07}
{Telleschi}, A., {G{\"u}del}, M., {Briggs}, K.~R., {et~al.} 2007, \aap, 468,
  541

\bibitem[{{Tsujimoto} {et~al.}(2004){Tsujimoto}, {Koyama}, {Kobayashi},
  {Saito}, {Tsuboi}, \& {Chandler}}]{tsu04}
{Tsujimoto}, M., {Koyama}, K., {Kobayashi}, N., {et~al.} 2004, \pasj, 56, 341

\bibitem[{{Ustamujic} {et~al.}(2016){Ustamujic}, {Orlando}, {Bonito}, {Miceli},
  {G{\'o}mez de Castro}, \& {L{\'o}pez-Santiago}}]{ust16}
{Ustamujic}, S., {Orlando}, S., {Bonito}, R., {et~al.} 2016, \aap, 596, A99

\bibitem[{{Zanni} \& {Ferreira}(2013)}]{zan13}
{Zanni}, C. \& {Ferreira}, J. 2013, \aap, 550, A99

\end{thebibliography}


\begin{appendix}
\section{Chandra observations of DG Tau}

   We studied the \textit{Chandra}/ACIS-S data set of DG Tau performed 
   in January 2010 (PI G{\"u}del; ObsID 11009, 11010 and 11011; 
   $t_{\mathrm{exp}}=120$~ks each observation), 
   centered at (04:27:04.80, +~26:06:16.90) (FK5).
   We reprocessed all the data in homogeneous way, using the latest CIAO 4.9 
   package. We studied the data individually and also as a unique archive with 
   $t_{\mathrm{exp}}=360$~ks, reprojecting the set of observations to a common 
   tangent point and creating a merged event file using the CIAO tools. We note 
   that all three data products correspond to observations done in the same week.
   We filtered the data in energy to study separately the hard and the soft 
   component. We chose the 0.5–1.0~keV band for the soft component and 
   1.5–7.3~keV for the hard one \cite[as in][]{gud11}. We explored different values 
   for the soft component band associated with the jet, e.g. 0.45–1.1~keV and 
   0.5–1.5~keV, obtaining almost identical results in all the cases. 
   Events were extracted for all observations from regions around the source 
   and the background, near the position of DG Tau (4:27:04.698, + 26:06:16.31).
   The images have been analyzed using the DS9 tool to study the morphology 
   of the sources, including the offset between the hard and soft component, 
   both at native \textit{Chandra}/ACIS spatial resolution and improving it by 
   performing the sub-pixel event-repositioning algorithm EDSER that can be 
   applied to \textit{Chandra} images to refine the event positions \citep{lij04}.
   We did not find a statistically significant offset between the two components.
   After applying the EDSER algorithm, the images can be resampled at 
   0.25\arcsec pixel size, obtaining images with one-half of the native ACIS 
   pixel scale (see left panel in Fig.~\ref{DGTAU}). 
   The asymmetry of the \textit{Chandra} point spread function (PSF) 
   has been investigated using CIAO tools to create a region that 
   highlights the location of this artifact for the source and we checked if 
   this instrumental effect may influence the observed morphology of the 
   X-ray source. We found that the asymmetry of the PSF does not affect 
   our images on scales larger than 1\arcsec, therefore, the elongated 
   structure visible in the images is not an artifact of the instrument. Finally, 
   we obtained the proper PSF for the data using CIAO and HEASOFT tools.

\end{appendix}


\end{document}